# Six Decades Post-Discovery of Taylor's Power Law: From Ecological and Statistical Universality, Through Prime Number Distributions and Tipping-Point Signals, to Heterogeneity and Stability of Complex Networks

Zhanshan (Sam) Ma[1,2,3]*　　　　R. A. J. Taylor[4]

[1]Faculty of Arts and Sciences, Harvard Forest
Harvard University
Cambridge, MA, 20138, USA

[2]Computational Biology and Medical Ecology Lab
Kunming Institute of Zoology
Chinese Academy of Sciences, Kunming, China

[3]Microbiome Medicine and Advanced AI Technology, Kunming, China
*Correspondence: ma@vandals.uidaho.edu

[4]Department of Entomology
Ohio Agricultural Research and Development Center
The Ohio State University, Ohio, 44691, USA.

## Abstract

First discovered by L. R. Taylor (1961, Nature), Taylor's Power Law (TPL) correlates the mean ($M$) population abundances and the corresponding variances ($V$) across a set of insect populations using a power function ($V = aM^b$). TPL has demonstrated its 'universality' across numerous fields of sciences, social sciences, and humanities. This universality has inspired two main prongs of exploration: one from mathematicians and statisticians, who might instinctively respond with a convergence theorem similar to the central limit theorem of the Gaussian distribution, and another from biologists, ecologists, physicists, etc., who are more interested in potential underlying ecological or organizational mechanisms. Over the past six decades, TPL studies have produced a punctuated landscape with three relatively distinct periods (1960s-1980s; 1990s-2000s, and 2010s-2020s) across the two prongs of abstract (mathematics/statistics) and 'physical' worlds. Eight themes have been identified and reviewed on this landscape, including population spatial aggregation and ecological mechanisms, TPL and skewed statistical distributions, mathematical/statistical mechanisms of TPL, sample *vs*. population TPL, population stability, synchrony, and early warning signals for tipping points, TPL on complex networks, TPL in macrobiomes, and in microbiomes. Three future research directions including fostering reciprocal interactions between the two prongs, heterogeneity measuring, and exploration in the context of evolution. The significance of TPL research includes practically, population fluctuations captured by TPL are relevant for agriculture, forestry, fishery, wildlife-conservation, epidemiology, tumor heterogeneity, earthquakes, social inequality, stock illiquidity, financial stability, tipping point events, etc.; theoretically, TPL is one form of power laws, which are related to phase transitions, universality, scale-invariance, etc.

**Running Head**: Six Decades Post-Discovery of Taylor's Power Law

**Keywords**: Taylor's Power Law (TPL); Population Spatial Distribution and Aggregation; Population Fluctuation and Stability; Heterogeneity and Diversity; Early Warning Signal (EWS) and Tipping Point; Complex Science and Complex Network

# 1. Introduction

In a 1961 letter to *Nature*, "Aggregation, variance and the mean", disclosed the discovery of a power function relating the mean ($M$) and corresponding variance ($V$) of a set of samples (Taylor 1961), which subsequently became known as Taylor's power law (TPL). In this Introduction we outline the history and significance of this discovery which was based on biological, principally insect, population samples (Taylor 1961, 1984) but has subsequently been found to apply also to many types of non-biological samples (Taylor, 2019). TPL has been validated across a diverse range of organisms and non-organic entities, both spatially and temporally, as well as abstractly (*e.g.,* the distribution of prime numbers) and concretely (*e.g.,* distribution of galaxies). The practical implications of studying TPL are extensive, influencing sectors such as agriculture, forestry, fisheries, wildlife conservation, climate changes, and even the prediction of turning points and worldwide mortality rates of COVID-19. The theoretical implications of TPL are even more profound, with extensive investigations conducted in fields such as ecology, evolution, economics, sociology, computer science, physics, as well as mathematics and statistics, to name a few major disciplines. This section concludes by outlining the structure of the review, which is organized around two key dimensions: a chronological timeline encompassing three periods and research themes covering eight topics, along with a future perspective.

## 1.1. Discovery of Taylor's Power Law (TPL)

Taylor's power law is named after the late British ecologist, L. R. (Roy) Taylor (Taylor, 2007). In the *Nature* paper Taylor showed that the variance ($V$) of samples from a set of $n$ populations of the same species occurring at spatially separated sites with different population densities accelerates as the average population density increases. The mathematical relationship between *V* and *M* is:

$$V = aM^b, \qquad (1)$$

where *a* and *b* are TPL parameters. A simple log-linear transformation of Equation (1) results in:

$$\log(V) = A + b\log(M). \qquad (2)$$

where *A* =log(a). Equation (2) is frequently used to fit TPL to the *V-M* datasets sampled from natural field populations.

Professor Sir Richard Southwood first introduced the term "Taylor's power law" in his classic textbook: "Ecological Methods: With Particular Reference to the Study of Insect Populations" (Southwood, 1966; Southwood & Henderson, 2000). Since Taylor's 1961 publication, TPL has been tested and verified by hundreds, if not thousands, of field observations of insects, plants, animals, humans, and more recently, microbes. A previous claim (Eisler et al. 2008) suggested that H. F. Smith (1938) was the first to publish TPL. However, as Meng (2015) clarified, Smith (1938) investigated the relationship between crop yield and plot size used to measure yields, not the average yield/plot. Meng's distinction is noteworthy as it highlights a unique feature of TPL, which relates the variance and mean of the same variable (quantity) rather than relating two variables that do not belong to the mathematical moments of the same random variable. There are several similar power laws in ecology alone, such as the well-known species-area relationship (SAR), which relates the number of species (S) discovered in an area of size (A) in power function form, $S = cA^z$. In fact, what Smith (1938) investigated is more similar to the SAR relationship than TPL. The species-area relationship was first reported in the 19th century in relation to plant biogeography. TPL, SAR, and other power laws share some commonalities, which we discuss in a later section, but they also have fundamental distinctions.

From a purely statistical perspective, TPL essentially represents the relationship between the first and second order moments of a random variable (X). Their exact function relationship is:

$$V[X] = E[X^2] - (E[X])^2, \qquad (3)$$

where *X* is the random variable. This means that the second order moment or the variance (V) of a random variable (*X*) is equal to the mathematical expectation or the mean $M=E[X^2]$ of the $X^2$ (squared random variable *X*) minus the square of the mean of the random variable (*X*). This relationship appears deceptively simple; one might intuitively expect that a mathematical proof (derivation) can be easily obtained by a competent mathematician. One might wonder if Equation (3) can be approximated with a power function,

$$V[X] = E[X^2] - (E[X])^2 \propto f[E(X)] \propto (?) \, aE(X)^b \qquad (4)$$

where $E[\cdot]=M$ is the mathematical expectation or the first order moment and $V[\cdot]$ is the variance or the second order moment about the mean. However, theoretically, the last functional relationship is not necessarily true. This is because the relationship can depend on the statistical distribution. For example, when $X$ follows a Poisson distribution, $V[X]=M[X]$. If $X$ follows a negative binomial distribution, the $V$-$M$ relationship is: $V[X] = M[X] + M[X]^2/k$, where $k$ is a dispersion or aggregation parameter. Different statistical distributions may generate different $V$-$M$ relationships.

While TPL was initially fitted to the $V$-$M$ data of biological populations, the power function itself does not require that the data must be from biological populations. Indeed, the V-M power function relationship has found applications well beyond its initial domain, insect populations. TPL has been verified for virtually all kinds of organisms including viruses, bacteria, fungi, animals, plants, and humans. Furthermore, TPL has been found to be applicable to many non-organic objects such as earthquake magnitude, galaxy distribution, network packages in computer networks, and fitness in evolutionary computing (*e.g.*, Ma 2012a, 2012b, 2013). The applications are distributed across virtually all fields where complexity sciences (Box S1) were successfully applied. The overlap may not be coincidental, since power-law scaling is often an emergent property of complex systems. Therefore, a significant portion of this review is not confined to biology or ecology. Even though many examples are indeed drawn from ecology, the underlying principles and patterns are applicable to any complex systems where the $V$-$M$ power-law relationship holds.

Although TPL was initially identified through fitting spatial (cross-sectional) sampling data of biological populations, it is equally applicable to temporal (time-series) sampling data (Taylor & Woiwod 1980). In fact, beyond the realms of biology and ecology, many TPL studies have concentrated on temporal scaling, such as fluctuations in the stock market and network traffic in computer networks. Indeed, the applications of TPL across space (referred to as cross-sectional samples or ensemble fluctuation) or time (known as time-series samples or temporal fluctuations) are related to each other. This relationship is contingent on the method used to compute the mean and corresponding variance (Ma 2015). For example, Cai et al. (2016, 2018) examines Taylor's law of ensemble fluctuation scaling in Chinese stock liquidity data. Analyzing the daily mean and standard deviation revealed a power law relationship, demonstrating Taylor's law holds. The scaling exponent averaged 1.442 but fluctuated daily. Removing outlier days with few data points tightened the exponent distribution. The robust power law over multiple magnitudes provides clear evidence that stock illiquidity complies with Taylor's law of cross-sectional fluctuation scaling, adding to the understanding of scaling in financial markets (Cai et al. 2016, 2018). TPL is also applied to study weather and climate changes (*e.g.*, Tippett & Cohen 2016, 2020).

## 1.2. Significance of studying TPL

The practical implications of studying TPL may not be immediately clear, but they are far-reaching. In sectors such as agriculture, forestry, and fisheries, population fluctuations of crops, trees, and fish can directly impact human food, fiber, and timber supplies, leading to significant economic consequences (Cohen & Schuster. 2012). In wildlife conservation, particularly for endangered species, population fluctuations towards low density can increase extinction risk (Allee effect) and potentially cause a genetic diversity bottleneck with enduring evolutionary implications. The fluctuations of insect pests and plant pathogens are a primary concern in plant protection (Cohen & Schuster. 2012), which may explain why TPL was first discovered in entomology. TPL has also been successfully applied in predicting turning points and worldwide mortality rates of COVID-19 (Ma 2020a, 2021). As a universal scaling law, TPL can play a critical role in studying the heterogeneity and stability (Box S1) of complex systems and networks. In physics and complexity science (Box S1), concepts such as universality, scale-invariance, phase transitions, and self-organized criticality are often connected with power laws, including TPL. For a detailed introduction, see Font-Clos (2015), also see Box S1.

## 1.3. Organization of this review

Beginning with L. R. Taylor's letter to *Nature* (Taylor, 1961) and culminating with R. A. J. Taylor's (2019) comprehensive monograph "Taylor's power law: order and pattern in nature", there have been countless original research papers, reviews, and book chapters published. Due to the vast scope of the TPL literature, our focus will be limited to the works that best illustrate and support the structure of this review. We have structured this review around two key dimensions (prongs): chronological timeline and research themes, as illustrated in Table 1. The review spans three time periods, each examining two main research themes: abstract and concrete. We try to adhere to these orthogonal combinations, starting from the inception point of (1961, The Discovery), through three periods and projecting to possible future directions (Heterogeneity and Stability of Complex Systems). However, when orthogonality proves challenging, we will primarily follow the trajectory of the research themes. This structure creates a 'punctuated landscape,' drawing an analogy from the fitness landscape concept in evolutionary biology. It presents a landscape of eight distinct topics (themes), as illustrated in Fig 1. These topics are covered in eight sections, which are positioned between the introduction and perspective sections.

**Table 1.** Outline of the review "Six Decades Post-Discovery of Taylor's Power Law"

| Dimensions | Outline |
|---|---|
| Timeline | 1960s—1980s: From population spatial aggregation to its ecological mechanisms |
| | 1990s—2000s: From ecological mechanisms to mathematical/statistical interpretations |
| | 2010s—2020s: From statistical distributions through tipping points to TPL-networks |
| | Future: TPL as a key metric for measuring heterogeneity and stability in complex science |
| Theme-line (18670 Words) | 1. Introduction (1580 Words) |
| | 2. TPL, Population Spatial Aggregation and Ecological Mechanisms (1970 Words) |
| | 3. Correspondence between TPL and Skewed Statistical Distributions (1560) |
| | 4. Mathematical/Statistical Generations (Mechanisms) of TPL (1500) |
| | 5. Sample *vs*. Population TPL & Distribution of Prime Numbers (1600 Words) |
| | 6. TPL, Population Stability and Synchrony (960 Words) |
| | 7. TPL on Complex Networks (2900 Words) |
| | 8. TPL and Other Classic Power Laws in Macrobial Ecology (1600 Words) |
| | 9. TPL in Microbiome Ecology (3500 Words) |
| | 10. Perspectives (1500 Words) |

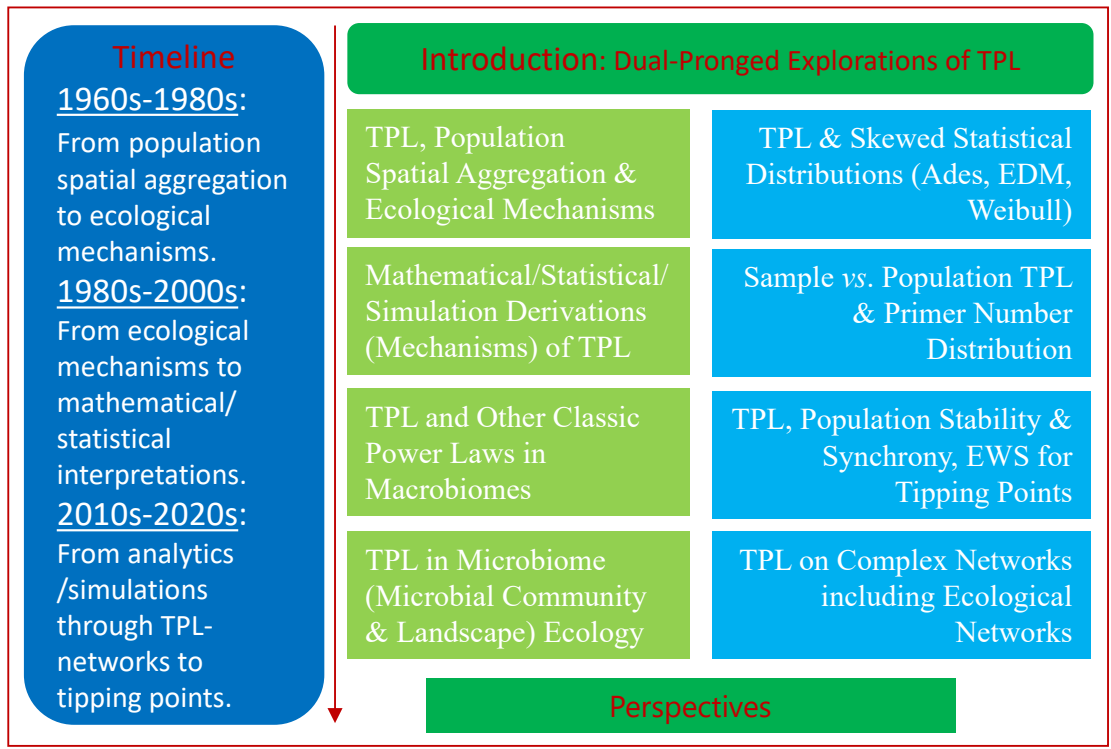

Fig 1. A 'punctuated landscape' representing the two-dimensional explorations of Taylor's Power Law (TPL) across eight thematic areas over past six decades

## 2. TPL, Population Spatial Aggregation and Ecological Mechanisms

This section reviews the early studies on using TPL in assessing the spatial aggregation (Box S1) of biological populations and the ecological mechanisms of TPL. The initial focus is on the spatial distribution of biological populations, particularly insect pests and plant pathogens, with Taylor's work, which involved two nationwide networks of traps sampling insect populations, one of which was later extended into Continental Europe. (Taylor 1961, 1984, 1986a, b). The section also discusses the debates between the Taylor and Iwao schools regarding which model is more suitable for assessing population spatial distribution patterns (Iwao 1968, 1972, 1977; Taylor 1984, 1986a, b). The practical applications of TPL/Iwao models, such as designing sampling schemes for assessing insect population density and performing data transformation methods, are also discussed (Taylor 1984, 2018, Ma 1989, 1992, Xu et al. 2017). The section further delves into the critiques against TPL, categorizing them into those who argue that TPL is an artifact of numbers rather than a genuine natural phenomenon, and those who believe that TPL has limited value or is even useless due to the lack of consensus on its underlying

mechanisms (Titmus 1983; Soberón & Loevinsohn 1987; Ma 1989, 1990,1991a, 1992; Shi et al. 2017).

In the first three decades of TPL, many of the studies were concentrated on the spatial distribution of biological populations, especially insect pests (Taylor 1984) and plant pathogens (Madden *et al.* 2018) of agricultural crops and livestock. Taylor engaged amateur moth enthusiasts to operate light traps throughout the UK and in 1964, he established a nationwide network of suction traps for surveying aphids. The network, known as the Rothamsted Insect Survey (https://repository.rothamsted.ac.uk/), monitors over 800 insect species, including aphids and moths. The aphid survey supports early-warning programs for mitigating both direct aphid damage and aphid-borne virus diseases. Both surveys provide invaluable datasets for a range of insect population studies, particularly for extensive testing of TPL (Taylor et al. 1978, 1979, 1980; Taylor 1984, 1986a, b). In today's terms, Taylor's work could be seen as a pioneering example of crowd science for big-data collection.

Interestingly, during the 1980s, in the wake of the Reforming and Opening-up policy, Chinese entomologists were motivated to resume active scientific research after a decade-long pause during the 'Cultural Revolution.' Many chose to study the spatial distribution of insect pests, possibly due to the relative ease of collecting cross-sectional survey data compared to observing population time-series data. Consequently, there was a substantial increase in TPL studies in Chinese literature, likely surpassing the volume found in English (e.g., Ma 1989, 1991a, 1991b, 1991c, 1992).

Also, between the 1960s and 1970s, population ecologist Iwao (1968, 1972, 1977) in Japan proposed a linear regression model ($M^*$-$M$) based on Lloyd's (1967) mean crowding concept (metric). This model established a linear relationship between mean crowding ($M^*$) and mean population abundance, represented as:

$$M^* = \alpha + \beta M, \qquad (5)$$

where α & β are the model parameters. The interpretations for α & β are similar to those for the parameters *a* and *b* of TPL.

Interestingly, there are parallel rules for using parameters TPL-b or Iwao-β for determining the spatial distribution patterns of biological populations. With b or β>1 indicating an aggregated (heterogeneous) population distribution, b or $\beta=1$ suggesting a random distribution, and b or $\beta<1$ indicating a uniform distribution. There were intense debates between the Taylor (1984, 1986a, b) and Iwao (1977) schools regarding which model is more suitable for assessing population spatial distribution patterns. In practice, both models seem to fit field observation data equally well, and in most instances, the conclusions from both models align.

In addition to determining the type of spatial distribution patterns (*see* Box S1) being aggregated, random, and uniform, two practical applications of TPL/Iwao models are designing sampling schemes for assessing population density (such as the optimum number of sampling units) and performing data transformation methods. The sampling design is a crucial part of monitoring insect population dynamics, which is necessary for decision-making in managing insect pest populations under the so-called economic tolerance level (ETL) (Taylor *et al*. 1998). Pest management measures are only economically beneficial when the insect pest population is above the ETL; below the ETL, no control measure is needed, saving the cost of control measures, avoiding pesticide pollution, while the economic loss from pest damage is tolerable. Data transformation is useful for performing statistical analysis when data violate certain necessary assumptions such as Gaussian normality and homogeneity of variances. These applications of TPL/Iwao models are critical for the so-called integrated pest management (IPM), which is essential for agricultural crop production and forest protection until today. TPL has been utilized to ascertain the minimum sequencing depth, specifically the minimal sequencing reads in amplicon sequencing (Ma 2020c). This application can be viewed as addressing a sampling problem in the field of metagenomics.

The debate over whether the *V-M* power law or the *M\*-M* linear model is more accurately fitted to field observation data, or more effectively interprets underlying ecological mechanisms, was intense. However, it's noteworthy that the power law model can also be applied with equal effectiveness to *M\*-M* observations, as demonstrated by Ma (1989). This equal fit does not resolve the complexities of ecological mechanisms and it simply reflects the flexibility of power law models. While both sides put forth persuasive arguments, it's crucial to acknowledge that both models are empirical as applied in early days of TPL. This trait is common to nearly all ecological models or laws, underscoring the idea that these models serve as tools for interpretation and prediction, rather than definitive depictions of ground ecological truth. The debates between the two groups gradually subsided following Iwao's passing in the late 1980s.

However, research work based on mean crowding continues to this day (e.g., Waters et al. 2014, Lang et al. 2017, Wade et al. 2018, Ma & Ellison 2018, 2019), albeit on a much smaller scale. For this reason, we will not delve into a comparison between the TPL/Iwao models in this review. Additionally, the critiques of TPL raised by third-party researchers, which we will briefly discuss below, are more extensive than those raised by Iwao.

In a different context, the concept of mean crowding, introduced by Lloyd (1967) and almost immediately harnessed by Iwao (1968, 1977) for modeling population aggregation in the context of comparative studies with TPL, was further explored by Ma & Ellison (2018, 2019) and has found applications beyond population aggregation. Utilizing the mean crowding concept, Ma & Ellison (2018, 2019) developed two related concepts: species dominance and community dominance, both within a unified mathematical framework and as functions of mean crowding. Species dominance measures the distance between a specific species and the "virtual gravity center" of the community of which it is a member. In this sense, species dominance is akin to species abundance, as both are species properties although they are essentially different properties. This similarity allows for the construction of a species dominance network (SDN), akin to a species co-occurrence network based on species abundances. Community dominance, on the other hand, is a function of the Gini-index in economics or the Simpson diversity index in ecology. In essence, dominance can be considered a measure of heterogeneity, as suggested by Shavit & Ellison (2021) and Ma & Ellison (2024).

The critiques against TPL can be categorized into two main categories: those who argue that TPL is an artifact of numbers rather than a genuine natural phenomenon, and those who believe that TPL has limited value or is even useless due to the lack of consensus on its underlying mechanisms, even if TPL does represent authentic phenomena. The most potent criticism of the first category is likely the critique on the fitting of TPL with log-linear transformation (regression), particularly the estimation difference between log-linear regression and non-linear optimization algorithms (Titmus 1983, Soberón & Loevinsohn 1987, Ma 1990, 1992, Shi et al. 2017). The second category of critiques focuses on the interpretations (mechanisms) of TPL for modeling the spatial/temporal distribution (aggregation) of biological populations in population ecology.

The initial hypothesis for TPL's mechanisms suggests the scaling parameter *b* is a species-specific evolutionary trait determined by density-dependent movement behavior, balancing migratory and sedentary tendencies due to resource competition (Taylor 1961, 1980, Taylor & Taylor 1977, 1979, 1983). Hanski (1980) proposed an alternative hypothesis: the multiplicative effect of reproductive success. Anderson et al. (1982) suggested random demographic events altering birth, death, and migration rates could lead to TPL, contradicting Taylor et al.'s (1983) proposition of non-random, species-specific migratory behavior. Most TPL criticisms are simulation-based. Perry (1988, 1994), Taylor & Perry (1988) used behavior-based simulation models generating TPL across various population densities, yielding species-specific responses. Hanski & Woiwod's (1993) simulations suggested TPL exponent (*b*) depended on the correlation between population's stochasticity (Box S1) and mean density. Downing (1986) presented data showing TPL exponent can vary within a species but be similar between different species, possibly reflecting both stochasticity and determinism's effects. Titmus (1983) argued against using the TPL scaling parameter to infer spatial pattern and animal behavior mechanisms due to potential spurious sampling influences. This criticism can be addressed with nonlinear optimization algorithms and/or extensive data collection (Ma 1990, 1992, Shi et al. 2017). Given the numerous field studies confirming good fittings, Titmus's (1983) and similar concerns are largely a moot issue.

Among the many simulation models for generating TPL, a few are particularly noteworthy. Taylor's (1981a, 1981b) delta model simultaneously generated both chaotic population dynamics and TPL. Perry (1994) also developed a simulation model to show that chaotic population dynamics could generate TPL, provided that the density-dependence effect is significant. Keeling (2000) constructed Markovian models for generating time-series simulations and produced TPL exponents. Kilpatrick and Ives (2003) demonstrated that species interactions could explain TPL for ecological time series data.

There was an initial debate about whether the TPL parameter (*b*) or exponent alone is important, or if the parameter (*a*) is also informative. In conventional discussions, the exponent (*b*) was used to identify the type of population spatial distribution, whether it is aggregated (*b*>1), random (*b*=1), or regular or uniform (*b*<1). Ma (1991a) introduced the concept of Population Aggregation Critical Density (PACD), which is a threshold of population density (abundance) at which the spatial distribution becomes random (V/M=1). The relationship between the type of population aggregation, population density, and PACD can differ depending on whether $b>1$, $b=1$, or $b<1$.

The PACD or $M_0$ was derived from TPL as follows:

$$M_0 = \exp\left\{\frac{\ln(a)}{(1-b)}\right\} \qquad (6)$$

where $a$ & $b$ are TPL parameters. When $b>1$, the population is classified as positively density-dependent (also known as inversely density dependent regulation), meaning the level of aggregation increases with the increase of population density. Specifically, when $M>M_0$, the population is aggregated, when $M=M_0$ it is random, and when $M<M_0$, the population is regular or uniform.

When $b<1$, the population is classified as negatively density-dependent (also known as density-dependent regulation), and the criteria for determining the aggregation are the opposite of the cases when $b>1$. Theoretically, when $b=1$, the population is density-independent, and the type of population aggregation is determined by the TPL-$a$ parameter alone: specifically, $a>1$, equals 1, and less than 1 should correspond to aggregated, random, and uniform distributions, respectively. In practice, it is more complicated because of the difficulty of sampling in the second quadrant [$\log(V)<1$, $\log(M)>1$] (See Fig. 1, Taylor & Woiwod 1982).

The insights gained from PACD ($M_0$) align with the clarification provided by Taylor, Lindquist and Shipp (1998) that the species-specificity of parameter b is not absolute and can be influenced by factors such as the age of insects (including stages like larvae and adults) and especially the sampled environment. This implies that parameter ($a$) can also be significant, and furthermore, their combinations and PACD can be crucial too. However, there are numerous instances where the TPL scaling parameter, denoted as 'b', appears to remain invariant. This is particularly evident in the spatial version of TPL, as exemplified in studies by Ramsayer et al. (2012), Li & Ma (2019) and Ma & Taylor (2020).

Moving beyond simulations and goodness-of-fittings of TPL, some researchers have attempted to develop mathematical derivations of TPL. Several statistical distributions, notably the negative binomial distribution, Neyman-A, Polya-Aeppli distribution, and Adès family distribution, have been comparatively studied with TPL on their roles in assessing population aggregation (Binns 1986, Kemp 1987, Perry and Taylor 1988). Arguably, the pinnacle of attempts to derive TPL from statistical distributions was reached through a series of efforts by Kendal (1995, 2004, 2015) with his scale-invariant exponential dispersion models (EDM), which is briefly reviewed below.

## 3. Congruence between TPL and Skewed Statistical Distributions

This section, consisting of four sub-sections, discusses the correspondence between TPL and certain skewed statistical distributions. The first subsection focuses on the Adès distribution, with Perry & Taylor (1986), investigating its application to insect population abundance data and finding that a family of Adès distributions can be fitted to satisfy TPL. The second subsection delves into the Tweedie distribution (Tweedie, 1984), with Kendal (2004) discovering that this class of scale-invariant exponential dispersion models (EDM) can generate TPL. Kendal & Jorgensen (2011) further found that TPL implies 1/$f$ noise when derived from sequential data, suggesting that TPL and 1/$f$ noise (Box S1) should be attributed to a universally applicable mathematical mechanism. The third subsection explores other statistical distributions, with Cohen & Xu (2015) discovering that TPL can emerge from any skewed probability distribution with four finite moments. De La Pena (2022) discusses a dynamic version of TPL for dependent samples using self-normalized expressions that involve Bernstein blocks. Taylor (1980) unified the dispersion models of insect populations with Weibull distribution, and Ma (1991$b$) found the congruence between Weibull distribution and TPL. The fourth subsection discuss TPL, Hurst scaling and Fractal dimensions.

### 3.1. Adès distribution

Perry & Taylor (1986) investigated the application of Adès distribution to insect population abundance data. Adès distribution is a family of distributions, named after J. M. Adès. The Adès distributions are continuous and are specified with an extra 'spike' of probability for the occurrence of a zero count. The Adès distributions are derived from a standard gamma variate, logarithmically transformed and then raised to a power. Assuming $X_i$ is a gamma distribution with the coefficient of variation $r$, and other parameter $\lambda$, $n$ is the number of gamma distributions (the datasets of frequency distribution), if $Y_i$ and $X_i$ is related by the following functions,

$$Y_i = (lnX_i)^{[2/(2-\tau)]}, \qquad X_i > 1, \tau < 2, \ i = 1 \ldots n. \qquad (7a)$$
$$Y_i = 0, \qquad X_i \leq 1. \qquad (7b)$$

then *Y* follows Adès distribution. The cumulative probability distribution function of Adès distribution is then,

$$Prob(Y_i \leq y_i) = I\{\lambda_i \exp\left[y_i^{[1-(\frac{\tau}{2})]}\right], r\} \tag{8}$$

where $I\{\bullet\}$ is the incomplete gamma integral. The probability of a zero value comes from the possibility that $X_i \leq 1$, when the logarithm would be negative and its power undefined. The exponent in the equation is positive when $\tau < 2$. For $\tau > 2$, the distribution requires a restriction because the distribution would have an infinite mean. This is achieved by restricting $Y_i$ to a maximum value, $Y_{max}$, and adjusting the remainder of the distribution to ensure that the probability sums to unity.

For any given value of $\tau$, a unique pair of (r, $\lambda_i$) values may be found which correspond to any desired values of the mean (*M*) and variance (*V*) of $Y_i$. Then Adès distribution can be fitted to satisfy a *V-M* power law relationship, with the parameters $\tau$ and *r* common to all distributions within a family, and one unique to each ($\lambda_i$) distribution. The parameters $\tau$ and *r* correspond to the parameters of TPL, with $\tau$ expected to be numerically equivalent to the slope of the power law (*b*) and *r* related to the intercept [ln(*a*)].

In summary, for a set of *n* animal populations of the same species occurring at different spatial or temporal points (usually different population densities), a family of Adès distributions with two shared parameters ($\tau$ & *r*) and one additional parameter ($\lambda_i$) for each population (site or timepoint) (a total of *n*+2 parameters) can be fitted to *n* datasets (populations). The same *n* datasets can be fitted to a TPL model (eqns. 1 & 2), and the parameter ($\tau$ & *r*) of the Adès distribution are then equivalent with the fitted TPL parameter *b* and ln(*a*).

## 3.2. Tweedie distributions

Kendal (2004) found that a class of scale-invariant exponential dispersion models (EDMs), known as Tweedie distributions, can generate TPL. He further discovered that one of the Tweedie distributions, the Poisson Gamma (PG) distribution, can be mapped to TPL with 1<*b*<2, under the assumption that each quadrat contains a random (Poisson-distributed) number of clusters, which would in turn contain a Gamma-distributed number of individuals. The Gamma-distributed clusters could presumably be determined by stochastic birth, death, and immigration processes. Indeed, in the population ecology of plants and animals, the majority of the *b*-values range between 1 and 2. Tweedie distributions comprise seven types, classified based on the magnitude of the exponent p, which is equivalent to the scaling parameter (b) of TPL. The PG distribution is the scale-invariant sum of a random (Poisson) number of gamma distributions. Some familiar distributions including Gaussian distribution ($p$=0), Poisson distribution ($p$=1), Gamma distribution ($p$=2), Inverse Gaussian distribution ($p$=3) are special cases of the Tweedie distribution family. Therefore, the Tweedie distribution family covers the limiting points of many other models, including the Gaussian distribution. Moreover, similar to the Gaussian distribution, there is a Tweedie version of the convergence theorem, akin to the central limit theorem of the Gaussian distribution. Since the Gaussian distribution is a trivial case of the Tweedie distribution family, it can be expected that the Tweedie family may play a significant role in describing both Gaussian and non-Gaussian processes (data).

In the context of observing time-series phenomena, the term frequency $f$ is used to denote the number of events per unit time sequence. This concept can be extended to non-temporal sequences, where frequency is now defined as the number of events observed per unit measure of the discrete series. The *1/f* noise (Box S1) refers to power spectra that exhibit an approximately inverse dependence on frequency and, similar to Taylor's Power Law (TPL), has been identified in numerous physical, biological, and socioeconomic systems. A modern paradigm for *1/f* noises is self-organized criticality. Kendal & Jorgensen (2011) discovered that TPL implies *1/f* noise when TPL is derived from sequential data using the method of expanding bins. Moreover, both TPL and *1/f* noise are central limit-like effects (convergence foci) of the class of Tweedie exponential dispersion models. They found that virtually any statistical model designed to generate TPL should converge to a Tweedie distribution. The Tweedie convergence theorem provides a unified mathematical explanation for the origin of TPL and *1/f* noise, found in a broad range of biological, physical, and mathematical processes. In other words, TPL and *1/f* noise should be attributed to a universally applicable mathematical mechanism rather than the *ad hoc* behavioral, biological, or physical processes that have traditionally been the focus of studies on TPL.

The Exponential Dispersion Model (EDM) characterizes error distributions for generalized linear models and can be employed to analyze a broad range of non-Gaussian distributions. There are two categories of EDMs, additive and reproductive, which can be transformed into each other through a duality transformation. Almost any statistical model or simulation designed to reproduce Taylor's Power Law (TPL) would mathematically converge to a Tweedie distribution, as a result of the Tweedie convergence theorem.

In conclusion, research on Taylor's Power Law (TPL) encompasses multiple aspects. Initially, TPL was employed to quantify the spatial distribution (aggregation) of single-species populations in their natural habitats. The method of expanding bins, used to evaluate sequential data, uncovers the consequence of spectral density determination of $1/f$ noise from TPL. The Tweedie Exponential Dispersion Models (EDMs) are found to be applicable to both approaches to TPL analyses. Therefore, the universality of TPL stems from the universality of the Tweedie models, which is in turn based on the Tweedie convergence theorem. TPL and $1/f$ noise are simply the general convergence behavior of non-Gaussian systems, with the Tweedie Poisson Gamma (PG) distribution being one of the mathematical convergence foci. Just as the central limit theorem validates a primary role for the Gaussian distribution in classic statistics, the Tweedie convergence theorem should establish a universal and fundamental role for the Tweedie distributions, considering their successful explanation of the apparent ubiquity and generality of both TPL and $1/f$ noise within many complex systems across natural and social sciences.

### 3.3. Other statistical distributions

Taylor (1978, 1980) unified insect dispersal models using the Weibull distribution in the exponential equation $\ln Y = a + bX^c$, where $X$ is insect recapture distance, $Y$ is catch density, and $c$ (Weibull exponent) shapes the dispersal curve. By optimizing $c$ (range: ±4) to maximize $r^2$, minimize residuals, and improve F-ratios, this framework adapts to diverse datasets. Taylor's approach, known as Taylor's exponential dispersal model (TEDM), demonstrates the Weibull function's flexibility in modeling insect dispersal, balancing generality and empirical precision. It remains influential for ecological studies of spatial population dynamics (Taylor 1978, 1980).

The relationship between TEDM and TPL lies in their mutual description of aggregation phenomena. When the TEDM shows fat-tailed distributions ($c < 1$), indicating some insects travel much farther than others, this corresponds to the aggregated populations ($b > 1$) described by TPL. Both models exhibit special behavior approaching random dispersion, though differently: TPL becomes linear ($b=1$) while losing aggregation discrimination, whereas the dispersal model's residuals become asymmetric as $c \to 0$, requiring modified interpretation.

Combined together, they reveal fundamental insights about the links between insect population dynamics and spatial distributions. Both TEDM and TOL complement each other in explaining insect spatial patterns at different scales. The dispersal model focuses on point-to-point density decay, while TPL examines variance patterns across sampling areas. Together, they form a powerful lens for understanding how nonlinear scaling governs insect distributions in nature. Taylor's work established that both dispersal behavior and local population interactions create these consistent, measurable patterns of aggregation that follow power-law relationships.

It was found that the Weibull distribution, in comparison to the negative binomial distribution – arguably the most frequently utilized statistical distribution for fitting insect populations – not only fits equally well to population abundances, but also possesses a mean and variance that empirically satisfy TPL, as demonstrated by Ma (1991a, b).

Cohen & Xu (2015) analytically demonstrated that from any skewed probability distribution with four finite moments, if independently and identically distributed samples are taken in blocks (not necessarily of equal size), then TPL can emerge with an exponent value between 1 and 2, but not greater than 2. They stated that their analytical derivation is only one of the possible mechanisms underlying TPL. The importance of their findings seems to further reinforce the universal nature of TPL—akin to the laws of large numbers and the central limit theorem in classic mathematical statistics regarding Gaussian distribution, even though no specific physical, biological, technological, or behavioral mechanism can explain all instances of TPL. This finding has been repeatedly confirmed in the literature of TPL, that is, TPL can be a mathematical property and the interpretations (generating mechanisms) of its scaling parameter can be problem-dependent, possibly very different in different complex systems, implying that TPL is an emergent property of some complex systems.

De La Pena (2022) discusses TPL, focusing on stationary time series, and develop different Taylor exponents in this context. The study introduces a dynamic version of TPL for dependent samples using self-normalized expressions that involve Bernstein blocks. The paper proves a central limit theorem under either weak dependence or strong mixing assumptions for the marginal process. Unlike the traditional framework where the limit behavior involves the marginal variance, this paper's limit behavior involves a series of covariances. The authors also provide an asymptotic result for a method to check if the dynamic TPL applies in empirical studies. This work appears to present a more rigorous formulation of concepts initially explored by Mandelbrot and Wallis (1969a, 1969b), which we briefly examine below

## 3.4 Taylor's power law, Hurst scaling and Fractal dimensions

A method for analyzing time series developed by Hurst *et al.* (1965) analyzed measurements of the water level of the River Nile. The method relates the rate at which autocorrelations in a time series, $\{x_t\}$ of length *T*, decrease as the lag or range, *t*, between pairs of values increases. Mandelbrot &Wallis (1969a, b) theoretically justified and extended Hurst's equation:

$$\mathrm{E}\left[\frac{R(t,s)}{S(t,s)}\right] \propto t^H \quad (9)$$

where E[·] denotes the expected value or mean of the range, $R(t,s)$ rescaled by the standard deviation, $S(t, s)$, both computed over the partial range from $t+1$ to $t + s$ (rather than to the total sample from 1 to *T*) for a selection of times, $2 < t < T/2$. The exponent, *H*, is the Hurst constant, so-named by Mandelbrot & Wallis (1968). The range, $R(t, s)$, is calculated from the difference between the maxima and minima $[\mathrm{Max}(z_t) - \mathrm{Min}(z_t)]$ of the accumulated sums of the observations, $\{x_t\}$, from $t+1$ to $t + s$, of the mean adjusted series $z_t = \sum_{t=1}^{s}(x_t - \bar{x}_t)$. The standard deviation of the partial time series, $S(t,s) = \sqrt{\frac{1}{s}\sum_{t=1}^{s}(x_t - \bar{x}_t)}$.

The Hurst exponent *H* is an index of long-range correlation. A value of $0.5<H<1$ indicates a time series with long-term positive autocorrelation, leading to longish periods of highs or lows. Series with $0<H<0.5$ have short-range negative autocorrelation with a high frequency of highs followed by lows and vice versa. $H = 0.5$ describes a completely uncorrelated series, but this true only in the long run. In the short run positive or negative autocorrelations with short time lags can occur but decay exponentially quickly to zero. Decay times for series with $H \neq 0.5$ is much longer. Mandelbrot & Wallis' (1969) analyzed 848 annual measurements of the height of the Nile at Cairo and estimated $H = 0.91$. This high value of *H* indicates that drought years are most likely followed by drought years and flood years by more flood years, leading Mandelbrot & Wallis (1968) to call sustained periods of lows and highs the Joseph and Noah effect. Other rivers have more modest values of *H* ($H = 0.55$ for the Rhine and $H = 0.69$ for the Loire, both based on monthly river flow).

Mandelbrot (1977) relates Hurst's *H* to the Hausdorff fractal dimension $D_H$ by

$$D_H = (2 - H). \tag{10}$$

The significance of this identity follows from the work of Zoltán Eisler and colleagues (Eisler & Kertész, 2006; 2007; Eisler *et al.*, 2008) on fluctuation scaling, physicist's name for Taylor' power law (TPL).

Eisler *et al.* (2008) developed a formal mathematical structure to account for fluctuation scaling in both temporal and ensemble scaling averaged over a collection of simultaneous samples (forming an ensemble set) or through time (a spatial set). Eisler *et al.* defined an entity they called "Activity" indexed by the subscript *i* during a specified time period $\Delta t$ ($A_i^{\Delta t}$). The activity could be the number and/or size of packets flowing through the *i*-th Internet router, the number and/or size of trades of the *i*-th stock on an exchange, or the number of moths caught at the *i*-th light trap or at a single trap on the *i*-th night. This collection of qualitatively different entities was a powerful way of studying quantitative scaling, but it led them to a misunderstanding of the nature of Fairfield Smiths (1938) power law which is closer to the number-area power law (considered later) than TPL. Also misunderstood was the nature of the data in Taylor (1961) by assuming the samples comprising the means and variances used in the power law plots came from different sized areas, instead of different sized population densities in equal sized areas.

By referring to sample sites as nodes as though considering networks and considering the activity at the *i*-th sample site or node as the sum of events (stock transactions, packets of information, number of animals or plants in time interval $\delta t = [t, t+\Delta t)$ they allowed for transactions and packets at site *i* to be of different sizes which they call "contributions" ($C_{i,j}$) numbered $j = 1$ to $N_i$. The activity in interval $\Delta t$ at time *t* at site *i* is

$$A_i(t) = \sum_{j=1}^{N_i(t)} C_{i,j}(t) \tag{11}$$

Assuming the variance of $A_i(\delta t)$ exists (which it always will in practice), it is given by the usual equation: $V[A_i(t)] = E[A_i(t)^2] - E^2[A_i(t)]$. The importance of Eisler *et al.*'s study is a proof for the variance of $A_i(t)$:

$$V[A_i(t)] = V[C_i(t)] \cdot E[N_i] + V[N_i] \cdot E^2[C_i(t)] \tag{12}$$

Eqn. (12) says the variance of all activity at a node or site is the product of the variance of the contributions and the expected number of events plus the product of the variance of the number of events and the expected contributions squared. By specifying that the number of events ($N_i$) is temporally and/or spatially long-term correlated in the sense of Hurst scaling, they found that the variance of the number of the contributions (events) at *i*, given *N*, $V(C|N)$ is proportional to the number of events raised to the power of $2H$:

$$V(C|N) = \mathrm{E}[(C - E[C])^2] \propto N^{2H} \tag{13}$$

Thus, the variance of the contributions to the activity at the *i-th* node or sample site in the interval $\Delta t$ is proportional to the number of events in the interval to the $2H$ power. If the $N_i$ are uncorrelated, then $H = \frac{1}{2}$ and variance is proportional to *N*; if they are long-range correlated $H > \frac{1}{2}$ and variance is proportional to a possibly fractional power of *N*.

In the context of organismal sampling, the variance of the contributions is the variance of the number caught (*N*) which we use to estimate the density, *M*. In their formal system applied to ecological sampling, Eisler *et al.* (2008) equate *b* and $2H$, leading to the following realization:

$$b = 2H = (4 - 2D_H). \tag{14}$$

which suggests that *b* may measure some attribute of a fractal structure. Though not described by Hausdorff's dimension, Mandelbrot recognized points in space which he called "dusts", as having non-integer dimension and thus fractal. The members of biological populations can be considered as being represented as points in space and *b* describes their degree of clustering or dimensionality.

Eisler *et al.*'s formalism simultaneously provides for a dependence of *b* on the size of the window $\Delta t$. For a fixed signal, in the presence of long-range temporal correlations, the variance can also grow by changing $\Delta t$. If the activity time series are long term correlated with Hurst exponents $H_i$ at node *i*, the window dependent variance at node *i* is

$$V_i(\Delta t) = E\{A_i - E^2[A_i]\}^2 \propto \Delta t^{H_i}$$

In this form, the time window size $\Delta t$ is the scaling variable instead of the number of constituents, *N*, and *V* increases with $E[A]^b$. $E[A_i]$ and $\Delta t$ are analogous and consequently

$$\Delta t^{2H} \propto E[A_i]^{b_{\Delta t}}. \tag{15}$$

The slope of this function is:

$$\frac{db_{\Delta t}}{d\log(\Delta t)} \approx \gamma \tag{16}$$

Eisler & Kertész (2006) identify several behaviours according to the value of γ. If γ = 0, the degree of temporal correlations measured by the Hurst exponent is the same at all nodes/sites and *b* is independent of $\Delta t$, and if γ > 0, the Hurst exponent at the *i*-th node, $H_i = H^* + \log(E[A_i])$ and $b_{\Delta t} = b^* + \log(\Delta t)$. Depending on the range of $\Delta t$ $\gamma$ can take two values in different ranges of $\Delta t$ with nodes having separate Hurst exponents. This latter condition is recognizable as the situation where TPL intersects the Poisson (ref to later).

# 4. Mathematical/Statistical Generating Mechanisms of TPL

In this section, we examine the mathematical and statistical foundations of Taylor's Power Law (TPL) across multiple disciplines. We first analyze how Cohen and colleagues (2012, 2013) applied TPL to population growth models, demonstrating how population synchrony shapes variance-mean relationships. Next, we discuss Jiang and coworkers' (2014) investigation of the temporal dynamics in TPL's scaling exponent, which revealed its context-dependent variability. We then evaluate Stumpf and Porter's (2012) critical assessment of power laws in natural systems, particularly their finding that many reported power laws lack rigorous statistical validation. Importantly, we consider Xiao and colleagues' (2015) provocative argument that TPL patterns may stem from numerical artifacts rather than biological mechanisms. Finally, we highlight Döring and collaborators' (2015) work drawing parallels between TPL and variance-mean relationships in agricultural systems.

In addition to analyzing the congruence between TPL and *V-M* relationship in highly skewed statistical distributions such as Adès and EDMs in the previous section, there is extensive literature on deriving TPL from population growth (dynamics) models. Competition is expected to reduce temporal heterogeneity and thus lower the temporal TPL slope. However, Ramseyer et al.'s (2012) experiment with two species of bacteria found that TPL ensemble *b* was unaffected by competition. Cohen et al. (2013) discovered that the Lewontin-Cohen model (Lewontin & Cohen 1969) for stochastic population dynamics can produce a spatial TPL in the limit of large time scales and provides an explicit, exact interpretation of the TPL parameters. The Lewontin-Cohen model is a geometric random walk model (different from the ordinary additive random walk model), which assumes that the combined effect of any demographic and environmental stochasticity is adequately captured by its multiplicative factor.

Notably, unlike the majority of studies on TPL mechanisms, Cohen (2013) conclusion was based on analytic derivation, rather than computer simulation, which lends theoretical solidity to their findings. However, unlike mathematical models in physics, the Lewontin-Cohen model itself (Lewontin & Cohen 1969) contains some ecological assumptions that may still be subject to experimental tests. To demonstrate this, Cohen (2013) analyzed the spatial heterogeneity (variability) and temporal dynamics of trees using long-term forest census data over 75 years.

According to Cohen (2014a, 2014b), theoretically, *b* can be variable and may even exhibit complex abrupt dynamic changes. This was empirically demonstrated by Taylor et al. (1998), who observed that gradual environmental transitions – particularly when samplers moved from open air into plant canopies – produced measurable disruptions in sampling consistency that altered *b* values. Cohen (2014a, 2014b) derived an abrupt transition in the scaling parameter (*b*) from a population growth model with a stochastic environment represented by a two-state, discrete-time Markov chain (one state of the environment leading to population growth and another leading to population decrease). He demonstrated an abrupt singularity of *b* in response to smoothly changing environments. Near the singularity, *b* could increase towards infinity, followed by a jump to negative infinity, and then return to 2 when beyond where the singularity occurs. Cohen's (2014a, 2014b) work is likely the first finding of possible sudden changes of TPL-*b* associated with gradual changes in the environment, essentially a tipping point (singularity) associated with phase transition of critical phenomena (Box S1).

Jiang *et al.* (2014) used a simple linear birth-death process model, in which each individual in a population has a probability $\lambda \Delta t$ of producing one offspring and a probability of $\mu \Delta t$ of dying in each small interval of time ($\Delta t$). The difference $r = (\lambda - \mu)$ is the intrinsic rate of growth per individual of the population. The expected population density at time *t* of a population is described with simple exponential model:

$$E[N(t)] = N_0 Exp(rt). \tag{17}$$

The variance is derived as:

$$Var[N(t)] = N_0 \left[\frac{\lambda+\mu}{\lambda-\mu}\right] \exp[(\lambda - \mu) t][\exp(\lambda - \mu) t - 1], \text{ if } (\lambda \neq \mu), \tag{18a}$$

and

$$Var[N(t)] = 2N_0 \mu t, \text{ if } (\lambda = \mu). \tag{18b}$$

The model was first developed by Pielou (1977), the probabilities of births and deaths are density-independent. When time goes to infinity, the population density goes either infinity or extinct (zero). Cohen (2014b) showed that for the linear birth-death process, if $\lambda > \mu$, then when $t \to \infty$, the process generates TPL $b=2$. If $\lambda < \mu$, then when $t \to \infty$, the process generates TPL $b=1$. If $\lambda = \mu$, then when $b$ is not defined. Jiang *et al* (2014) filled the gap left by Cohen (2014b) by estimating the scaling parameter $b$ for finite time periods ($t$) under different initial population density ($N_0$).

Jiang *et al.* (2014) defined the "transient" value of $b=b(t)$ at a finite time $t$ as the derivative in the form of:

$$b(t) = \frac{d\{lnVar[N(t)]\}}{d\{lnE[N(t)]\}}. \tag{19}$$

The transient $b(t)$ can be derived exactly via limits of $\Delta t \to 0$, as $b(t) = \frac{2-\exp[-(\lambda-\mu)t]}{1-\exp[-(\lambda-\mu)t]}$. Obviously, $b(t)$ is the slope of tangent lines at time $t$, and it has singularity points, which predict the abrupt changes of $b(t)$. $b(t)$ can be positive, negative, or even infinity, depending on the values of $\lambda$, $\mu$, $t$, and $N_0$. Jiang et al. (2014) obtained three scenarios: $b$ with singularity nearby 2, $b \approx 1$ and $b$ is U-shaped function of $\lambda$. Despite the simplicity of the population model Jiang et al (2014) adopted, in reality, the model is hardly realistic given its lack of non-linear density dependence and the model is essentially a stochastic version of the Malthus exponential population growth model. However, the behavior of the transient scaling parameter ($b$) is sufficiently rich to remind us of the complexity of TPL scaling under influences of population ages (population is sampled at different time $t$) and initial population size. The discovery of a singularity in $b$ at $\lambda \approx \mu$ (when net growth is near zero) suggests that when the average population density is stable, the population variance can continue to increase due to random births and deaths. These findings underscored the importance of TPL-$b$ in interpreting both population temporal stability and spatial heterogeneity under stochastic environments. Fujiwara & Cohen (2015) further applied Jiang et al. (2014) metric of transient (local) exponent to investigate whether or not it can exhibit abrupt changes in stage-structured, density-dependent fishery models with environmental stochasticity only. Stage-dependent models are usually necessary for modeling population dynamics with complex life histories such as insects and fish.

Döring *et al.* (2015) showed that TPL is an effective method for illustrating the relationship between the variance and mean of crop yields. They found that TPL is often applicable to various factors such as crop species, varieties, environments, or countries, but not always.

Interestingly, the application of the power law relationship to depict crop yield predates the establishment of TPL itself, before L. R. Taylor's discovery of TPL (Taylor 1961). H Fairfield Smith (1938) (Cited in Eisler et al. 2008) was the first to publish a similar equation for crop yields, but Fairfield Smith's (1938) work did not relate variance to the mean of crop yield, instead, related to the plot size used to measure yield, which is somewhat similar to species-area relationship (SAR), as mentioned previously. The variance of crop yields is crucial as it indicates the stability of food supplies. Instead of using the TPL scaling parameter to gauge yield stability, Döring et al. (2015) proposed using power law residuals—calculated as the residuals from the linear regression of log(V) against log(M)—as a stability metric. They demonstrated that this POLAR (Power Law Residual) measure offers superior performance compared to conventional stability indices. However, residual's sensitivity to data quality may limit its utility for mechanistic ecological interpretations as demonstrated in Stumpf and Porter (2012).

Stumpf and Porter (2012) discuss the prevalence of power laws in various fields, noting the often insufficient statistical support for these laws. Their discussion primarily focuses on power-law statistical distributions, but their insights are also relevant to Taylor's Power Law (TPL). Both the univariate statistical distribution of power law and the bivariate TPL are frequently used as models to describe highly skewed, non-random complex systems, often in disequilibrium states. They highlight that even when a power law statistically fits the data, a clear mechanistic explanation is often lacking. They cite allometric scaling as one of the few power laws with strong statistical and mechanistic support. The authors caution that noise in data can distort apparent power law relationships and that power laws can emerge from statistical phenomena without a specific underlying mechanism. They advocate for a more critical approach to claiming the existence of power laws, emphasizing the need for robust statistical and theoretical justification.

Xiao et al. (2015) suggested that Taylor's law may arise from numerical constraints on the total number of individuals and the number of groups they are distributed among, rather than specific ecological processes. They sample from the "feasible set" of all possible configurations of distributing individuals among groups to generate artificial mean-variance relationships. These relationships closely match empirical Taylor's laws, but cannot accurately predict variance values or individual relationships, suggesting additional ecological factors shape detailed patterns. The study offers a general statistical explanation for Taylor's law ubiquity, but not for specific details.

Although TPL can distinguish between life-system parameters generating unique TPL relationships (Park *et al*. 2013), the ability of TPL to recover life-system parameters is very limited. Samaniego *et al*. (2012) found they could reproduce field sampled TPLs with Markov chain population models incorporating life-history and spatial behaviour parameters in two closely related species of jay: piñon jays (*Gymnorhinus cyanocephalus*) and western scrub jays (*Aphelocoma californica*). Piñon jays are social, non-territorial and have a specialized diet while scrub jays are comparatively polyphagous and aggressively territorial. Both species occupy semiarid habitats that differ principally in elevation. Their field data were derived from the annual Breeding Bird Survey that follows fixed routes in southern Canada and the contiguous United States (BBS; Robbins et al., 1986). They modeled colonization-extinction processes using a 2-state Markov chain (Norris, 1998) to generates transitions between four states (persistence, continued absence, colonization, and extinction) derivable from the colonizations and extinctions observed on the BBS routes. By manipulating several life-system parameters, the model was tuned to mimic the variance mean slopes computed from the BBS field data. Their results supported the conclusions that difference in slope between the species was at least partly the result of behavioral differences. Although the model could mimic the TPL of the field data they could only recover the parameters from the field data by simulation. In short, life-system parameters can generate TPL parameters.

## 5. Sample *vs.* Population TPL, and Distribution of Prime Numbers

In this section we discuss two apparently independent topics: the distinction between sample and population TPL models and the distribution of prime numbers (Kendal & Jørgensen 2015; Giometto *et al*. 2015; Zhang *et al*. 2016). Zhang *et al*. (2016) propose using quantile regression to analyze log-periodic power law singularities in financial time series data, providing a tool to determine whether observed patterns of population abundance are dependent on the underlying population dynamics. Kendal & Jørgensen (2015) demonstrated that the distribution of prime number, measured in terms of the deviation $D(x)$ between Riemann prime counting function $R(x)$ and actual prime counting function $\pi(x)$, exhibits that the mean and variance of $D(x)$ follow TPL. The reason we include these two topics in the same section is that the TPL for prime number distribution is based on a sampling process whose results depend in part on the details of the sampling strategy. It turns out that different regions of integers sampled can produce different

TPL scaling parameters. In other words, there is not a universal scaling parameter for the prime number distribution.

## 5.1. Sample *vs.* Population TPL

Determining whether the observed macroecological patterns are due to statistical artefact rather than actual ecological mechanisms can be quite complex. The finite size of both ecosystems and sampling endeavours could potentially lead to statistical artefacts, even when the authentic ecological processes are in operation.

Given that TPL parameters must be practically estimated using a finite number of data points (samples), it is crucial to differentiate between the practical (empirical fitting) or sample TPL model and the theoretical population TPL model. The latter assumes an infinite number of samples for fitting or deriving analytically from theoretical population models. Consequently, one could argue that a TPL model estimated in practice may not accurately reflect the true pattern, even if TPL is theoretically applicable to the process (pattern) being studied. In other words, the practically fitted TPL model could be a statistical artefact. To address this problem, it is essential to identify the minimum number of samples or adequate sampling efforts needed to secure reliable sample TPL parameters that consistently approximate the population TPL parameters. It is important to note that the term "population" in this sub-section is used in a statistical context (*sensu* statistics), which differs from its biological usage as a "population" (a collection of all individuals) or *sensu* biology in other sections.

Giometto *et al*. (2015), drawing on multiplicative population growth models (sensu biology) in Markovian environments (Cohen 2014a, Theoretical Ecology; 2014b Theoretical Population Biology), developed generalized TPL models for both population (sensu biology) and sample versions. The model for population growth takes the following form:

$$N(t) = N_o \prod_{n=1}^{t} A_n \qquad (20)$$

where $N_0$ is the initial population density (*sensu* biology), $N(t)$ is the population at time ($t$), and $A_i$ is the multiplicative growth factors determined by a two-state Markov chain with state space {*f*, *h*}, for example, *f* for favorable environment in which reproduction occurs, and *h* for hostile environment in which reproduction cannot occur. They found that with this growth model, there is a power function relationship between the moments of the random variable N(*t*), *i.e.*, the generalized TPL model in the form of:

$$E[N^k(t)] = a_{jk} E[N^j(t)]^{b_{jk}}. \tag{21}$$

The GTPL (generalized TPL) holds asymptotically in *t* for any choice of *j* and *k* (orders of moments), both for population (sensu statistics) and for sample moments. When *j=1*, and *k=2*, the GTPL becomes the traditional TPL.

By applying large deviation theory techniques and finite sample size arguments, Giometto *et al*. (2015) showed two regimes ($t \gg \log R$ and $1 \ll t \ll \log R$) exist, where *R* is the number of finite number set of *R* independent realizations of the process obeying TPL. In the first regime ($t \gg \log R$), sample exponents inevitable tend to ($b \cong 2$) $or$ ($b_{jk} \cong \frac{k}{j}$) regardless of the underlying population exponent. In the second regime, ($1 \ll t \ll \log R$), sample exponents accurately approximate population exponents, which can be calculated analytically and may be different from *b=2*. As such, the sample exponents are predictively dependent on the number of samples (sampling efforts), and a sample exponent of *b=2* could potentially be a statistical artefact. If there's a discrepancy between the sample and population, the use of TPL could lead to unfavorable outcomes, such as failing to detect early warning signals (EWS) (Box S1) for tipping points (sudden changes in exponent values) or inaccurately assessing the stability (extinction risk) of endangered species' populations. The study by Giometto *et al.* (2015) provides a tool to determine whether observed patterns of population abundance are dependent on the underlying population dynamics.

Zhang et al. (2016) developed a quantile regression approach to detect financial bubbles using log-periodic power law singularities (LPPLS) models. Quantile regression, unlike standard regression that estimates average relationships, models how different percentiles (e.g., 10th, 50th, 90th) of the response variable depend on predictors, making it particularly robust for identifying bubble patterns across market conditions. The LPPLS model specifically captures super-exponential growth through a power law term, with exponent values between 0.1-0.9 signaling bubble conditions. By applying quantile regression across multiple time scales, they generate comprehensive bubble diagnostics less sensitive to market noise. Their DS LPPLS Confidence and Trust indicators aggregate these signals, showing superior performance to conventional methods in tests across 16 historical bubbles – providing earlier warnings (typically 3-6 months sooner) with fewer false positives (22% reduction).

While focused on finance, Zhang et al. (2016) approach offers three advantages for ecological early warning systems using Taylor's Power Law: (i) ability to detect different transition phases through distributional analysis, (ii) robust handling of noisy ecological data, and (iii) probabilistic confidence measures for stability thresholds. The method's multi-scale quantile framework could significantly enhance TPL's predictive power for ecosystem stability analysis.

## 5.2. The distribution of prime numbers

The goal of predicting the positions of prime numbers within the sequence of integers has long captivated the interest of mathematicians, engineers, and physicists, not only due to pure mathematical curiosity but also because of its practical applications in cryptography and other fields. The prime counting function, π(x), provides the actual count of prime numbers up to the positive real value x, taking the form of a step function that increments by 1 with each new prime number. In 1808, Legendre demonstrated through empirical evidence that π(x) could be approximated using the formula x/[log(x)−B], where B is a numerical constant. Gauss further refined the estimation of π(x) by employing the logarithmic integral (LI), which is defined on the positive real numbers x ≠ 1.

$$LI = \int_0^x dt/\ln(x). \tag{22}$$

The approximations prompted the conjecture of prime number theorem:

$$\pi(x) \sim x/\ln(x). \tag{23}$$

Bernhard Riemann worked out a more detailed formula for the approximation:

$$R(x) = \sum_{n=1}^{\infty} \left(\frac{\mu(n)}{n} LI(x)^{1/n}\right), \tag{24}$$

where $LI$ is the log-integral function, $\mu(n)$ is the Möbius function and $n$ is the integers. Which is defined as μ(n)=1 if the integer n=1, μ(n)=0 if n has one or more repeated prime factors, and μ(n)=(−1)$^k$ if n is a product of k distinct primes.

Kendal & Jørgensen (2015) used the deviation between $\pi(x)$ and $R(x)$, which is conjected to follow the following equation:

$$D(x) = R(x) - \pi(x) \sim \sum_\rho R(x^\rho) \tag{25}$$

to study the local behaviour of prime distribution. The summation is done over the nontrivial (complex) zeros of ρ of the Riemann zeta function. The above summation is conjectured to be equivalent to the Mertens function:

$$M(n) = \sum_{i=1}^{n} \mu(i) \tag{26}$$

where $\mu(i)$ is the Möbius function.

The deviations $D(x)$ exhibit chaotic traits and long-range correlations, hinting at a hidden structure. In fact, the positional irregularities (fluctuations) of prime numbers bear the hallmarks of $1/f$ noise and can be linked to the eigenvalues of random matrices used in quantum chaos. However, the reason why the distribution of prime numbers should show such traits remains elusive. Kendal & Jørgensen (2015) found that the deviations $D(x)$ empirically align with a probabilistic model marked by a power law relationship between the variance and the mean. This model falls into the category of scale-invariant, or Tweedie, exponential dispersion models (EDMs) discussed previously. They inferred that $1/f$ noise, Tweedie-EDM, self-organized criticality, and TPL models discussed previously should be applicable to describe the distribution of prime numbers (Kendal & Jørgensen 2015).

Kendal & Jørgensen (2015) estimated the absolute values of the prime number deviations |D(x)| for sequential, equal-sized enumerative bins that cover the integers. The mean and variance of the values of |D(x)| within each bin were estimated over the area of interest. This procedure was repeated for progressively larger bin sizes, enabling the construction TPL, which was verified by their experiments. Empirical cumulative distribution functions (CDFs) were also developed from the original sequences of |D(x)| and fitted to theoretical CDF from the Tweedie compound Poisson-gamma distribution, which intrinsically demonstrates TPL with scaling parameter between 1 and 2. For examples, Kendal & Jørgensen (2015) found that different regions of integers can produce different TPL-b values. For the first 50,000 integers, *b* was 1.83. The TPL for |D(x)| for $x = 10^6 \ldots 10^7$ in steps of $10^3$ produced *b*=1.66.

Kendal & Jørgensen (2015) concluded that the Riemann deviations $D(x)$, closely tied to prime number positions, appear to behave as Tweedie probability distributions, and to follow TPL, possibly due to a central limit-like effect. They emphasized that any model producing TPL must converge towards a Tweedie model and yield 1/*f* noise. Self-organized criticality, a hypothesis explaining 1/f noise and power law scaling behaviors, is often demonstrated through sandpile model simulations. However, these simulations also show a variance to mean power law and conform to the Tweedie distribution, suggesting that phenomena attributed to self-organized criticality could be due to Tweedie model convergence behavior. This implies that the irregular distribution of prime numbers could be another example of the TPL fluctuation scaling linked to this mathematical convergence behavior. Self-organized criticality is thought to occur in dynamical systems that naturally reach borderline unstable states without external manipulation. Yet, the 1/f noise seen in the Riemann deviations might be due to a mathematical convergence effect related to the central limit theorem. Further research could change our understanding of processes traditionally attributed to self-organized criticality (Kendal & Jørgensen, 2015).

## 6. TPL, Population Stability and Synchrony

This section delves into how synchrony, a phenomenon where time series of population densities in various spatial locations correlate over time, affects the linearity and slope of TPL (Reuman et al., 2017) and how small-scale spatial synchrony relates to temporal TPL (Ballantyne & Kerkhoff, 2005). We further consider how TPL can be used to evaluate population dynamics models by comparing simulation data based on a population dynamics model with empirical data (Cohen & Saitoh, 2016). The section concludes by highlighting the potential applications of exploring the relationship between TPL and synchrony, with possible consequences for resource management, conservation, and other sources of systemic risk.

Synchrony, also known as spatial synchrony or metapopulation synchrony, is the propensity for time series of population densities, observed in various spatial locations, to correlate over time. Synchrony operates within the metapopulation framework, though in practice, strictly defining the distributions or boundaries of a local population can be difficult. A particularly intriguing aspect of synchrony theory is the so-called "Moran effect," which refers to synchrony induced by simultaneous environmental drivers. For instance, it is broadly speculated and actively researched whether climate change has been modifying population synchrony in general.

Similar to TPL, synchrony may also reflect aggregation (Box S1), as the spatial extent of correlations among population time series is dictated by the geographic size of outbreaks. Therefore, synchrony and TPL are two interconnected fields. However, it should be noted that synchrony is the phenomenon, and TPL merely reflects the outcome of this phenomenon. Reuman *et al.* (2017) investigated two key questions regarding the relationship between synchrony and TPL. The first question was whether the presence and intensity of synchrony in population time series affect the validity of TPL, that is, the disruption or maintenance of TPL, and if so, in what way? The second question was, if the answer to the first question is affirmative, does the TPL scaling exponent (*b*) get influenced by synchrony, and if so, how?

Reuman et al. (2017), through extensive analytical, simulation, and empirical studies, discovered that the strength of synchrony can significantly impact the relationship between log (*V*) and log (*M*), where TPL is valid if the relationship is linear (log-linear in terms of *V vs. M*). Synchrony can disrupt this linearity, but more often, it preserves the linearity while changing the scaling exponent (*b*). Notably, a high level of synchrony typically reduces the TPL exponent. When synchrony increases from zero, the TPL slope (*b*) almost invariably drops sharply.

One systematic deviation from the general discovery is that when the marginal distributions of time series in various locations are identically and normally distributed, then a nonzero slope of TPL is unexpected, irrespective of the presence of synchrony. In simpler terms, log (*V*) and log (*M*) can be independent when the marginal distributions are identically and normally distributed. Reuman *et al*. (2017) even devised a method to partition exponent (*b*) to $b_{sync}$ and $b_{marg}$, which represent the portion attributable to synchrony and the portion due to the marginal distributions of time series, respectively.

Ballantyne & Kerkhoff (2005) and Eisler et al. (2008) explored the connection between small scale spatial synchrony and temporal TPL. They discovered that the type of synchrony they examined can generate a TPL with an exponent of *b*=1. TPL is also suggested to be crucial in evaluating population dynamics modeling. Cohen & Saitoh (2016) constructed temporal and spatial TPL, respectively, for the empirical time-series density of the Hokkaido gray-sided vole at 85 locations. They then compared the parameters of the fitted TPL model with the parameters of the TPL model fitted with simulation data based on a population dynamics model previously created for the same datasets, specifically the autoregressive Gompertz model. The comparative analysis of the same vole datasets using both TPL and Gompertz-based simulations enabled them to identify the unique effects of temporally density-dependent and temporally density-independent regulatory factors (which are likely spatially correlated or synchronized perturbations such as habitat heterogeneity). Clearly, a significant challenge is that both spatial and temporal effects are inseparable.

When modeling temporal fluctuations with TPL, the TPL temporal scaling parameter can vary across different spatial sites – it could be independent, dependent, or even synchronized. A method to categorize populations based on the TPL temporal scaling parameter, in conjunction with other environmental and intrinsic factors (like initial population abundances and growth rates), could be the Fuzzy clustering and Grey-system clustering algorithms (Ma 1991c). Notably, these algorithms do not necessitate precise quantifications on the states of population dynamics, such as equilibriums or outbreaks, as demonstrated by Ma (1991c).

In summary, the exploration of the relationship between TPL and synchrony carries significant potential applications due to the ubiquity of both synchrony and TPL, and the importance of investigating metapopulation dynamics, which may be as crucial, if not more so, than studying local population dynamics. The potential implications arising from the synchrony-TPL relationships could permeate various fields such as resource management, conservation, human demography, agricultural production, tornado outbreaks, stock markets, and systemic risk in financial systems. Specifically, the investigation into the synchrony-TPL and early warning signals (EWS) for tipping points (TP) (Box S1) could be of immense importance for potentially predicting black swan events.

## 7. TPL on Complex Networks

This section reviews Taylor's Power Law (TPL) applications across complex, ecological, and scale-free networks, integrating perspectives from information theory, graph theory, and network science (Box S1). Subsection 7.1 discusses the application of TPL on complex and computer networks, with de Menezes & Barabasi (2004) applying the power law in the form of $S \propto m^{\alpha}$, where $m$ represents the average flux, and $S$ is the dispersion measured as the standard deviation. Eisler & Kertész (2005, 2006) and Duch & Arenas (2006) further explore the universality of the scaling exponent. Subsection 7.2 introduces ecological network applications (Ma 2024), while Subsection 7.3 evaluates scale-free networks through network science and graph theory lenses, reconciling Barabási & Albert's (1999) foundational scale-free network findings with Broido & Clauset's (2019) empirical limitations and Holme's (2019) emphasis on heavy-tailed distribution identification over strict power law conformity (Box S1). Subsection 7.4 analyzes information theory integration, including Freitas et al.'s (2019) network entropy framework and the evolving relationship between random graph theory (Watts & Strogatz 1998; Barabási & Albert 1999) and modern network science, while noting persistent disciplinary gaps between graph theory and network science (Iñiguez et al. 2020) (Box S1).

## 7.1. Taylor's power law on complex networks and computer networks

de Menezes & Barabasi (2004) applied the power law to a complex network in the form of $S \propto m^\alpha$, where $m$ represents the average flux, and $S$ is the dispersion measured as the standard deviation. This is clearly a version of Taylor's power law with α = b/2. de Menezes & Barabasi (2004) discovered that a scaling exponent of α=1/2 encapsulates an endogenous behaviour dictated by the system's internal collective fluctuations, which originate from the randomness (Box S1) in the walkers' arrival and diffusion. Conversely, an α=1 exponent represents driven systems, where the fluctuations of individual nodes are primarily influenced by temporal changes in the external driving forces.

Eisler & Kertész (2005) utilized a version of temporal Taylor's power law, expressed as $S(i) \propto \langle m(i) \rangle^\alpha$, where $m(i)>0$ is the average activity of node ($i$) and $S(i)$ is the standard deviation of the activity at node ($i$). The universality class of $\alpha$=½ is indicative of equilibrium and the dominance of internal dynamics. The universality class of $\alpha$=1 is prevalent in systems with a strong driving force, where the dominant factor is typically externally imposed dynamics. They introduced an impact variable, using a random walk model, where the activity is the product of the number of visitors (akin to the network degree) at a node and their impact (similar to species abundance). If the impact (abundance) is heavily dependent on the node degree (connectivity) and the properties of the carrying network are broadly distributed, as in a scale-free network, they found non-universal scaling values. The exponent always surpasses the universal value of *1*, if the external drive is dominant. Eisler & Kertész (2006) further applied their model to examine the fluctuation scaling of stock values, and found values of α between ½ and 1. Intermediate values may have different origins and offer insights into the microscopic dynamics. Duch & Arenas (2006) identified a wide range of exponents between ½ and 1, which raises questions about the existence of universality classes.

## 7.2. Taylor's power law of ecological network (TPLoN)

Conceptually, let $G = \{Nodes, Links\}$ represent a species co-occurrence network constructed from the dataset of an OTU (operational taxonomic unit) or SAD (species abundance distribution) table. This table is typically a matrix, with each row consisting of the species abundances of all species in a community sample. Consequently, all rows of the matrix constitute a metacommunity under investigation. The network $G$ is essentially a graph, with nodes representing the set of species in the metacommunity and links representing the set of all co-occurrences (a proxy for interactions) between the species. Suitable correlation coefficients can be selected to construct the species co-occurrence network, with the choice depending on the research objective and data available. The species co-occurrence network, serving as a proxy for the species interaction network, operates on the metacommunity scale. The process of porting TPL onto the network (TPLoN) is also performed on the metacommunity scale, *i.e.*, applying the TPL extensions (TPLE) (Ma 2015) to the network. It's important to note that the process of building species co-occurrence networks or species interaction networks is independent of TPLoN. This is because TPLoN analysis, as described below, can be applied to any other ecological networks beyond correlation networks, and in fact, to any complex network.

Once the network is constructed, the TPLoN analysis can be distinguished as the following four steps, and the first two steps define and compute the so-termed species connectedness (SC) and weighted species connectedness (WSC), respectively. The WSC for each species (network node) is a vector of real values, and the third step is simply to compute the mean ($M$) and Variance ($V$) of the vector elements. The TPLE is then fitted by regressing $V$-$M$ series across species (network nodes). The TPLoN parameters can be harnessed to characterize the network heterogeneity of metacommunity.

The first key metric for implementing TPLoN analysis is to define and compute the connectedness of species,

$$C_i = (D_i/A_i), \tag{27}$$

where $D_i$ is the network degree of species (node) $i$, $A_i$ is mean abundance of species $i$ across all samples used to construct the network. In other words, the *species connectedness* ($C_i$) is the number of significant correlations (network links) per individual of species $i$. The motivation to define species connectedness is to capture the capacity (or tendency) of an individual of species ($i$) to interact (correlate) with individuals of other species in the network.

The second metric (also the step) is to weigh the species connectedness by species correlation coefficients, and the resulting weighted species connectedness (WSC) of species $i$ synthesizes the information from *degree distribution* and *species interaction strength* (*e.g.*, measured by species correlation coefficients). The WSC of species $i$ is defined as:

$$WSC[i] = C_i \times [R_{i1}, R_{i2}, R_{i3}, ... R_{iD_i}] \tag{28}$$

where species connectedness $C_i$ is defined previously, $R_{ij}$ stands for the species correlation coefficient between species $i$ and $j$, $j$=1, 2, … $D_i$, and $D_i$ is the degree of species $i$ defined as the number of direct links with its nearest neighbors. The WSC is a *vector* of the product of the species connectedness ($C_i$) and the correlation coefficients with its direct neighbors.

The third step in implementing the TPLoN is to define and compute the *mean* (*M*) and *variance* (*V*) of the vector elements for each species' WSC[*i*] vector. Consequently, there is a pair of *variance* (V) and *mean* (M) for each species or network node in the network, and a series of *V-M* pairs across all nodes of the network are obtained. The final is straightforward, fitting the TPL function ($V = aM^b$) with the previously defined and computed *V-M* pair series. The fitted TPLoN parameters (*b* and *a*) measure the scaling of network heterogeneity. A derived parameter, NHCT (network heterogeneity criticality threshold), is similar to the population aggregation critical density (PACD) previously defined as the density $M_o$ [Eqn. (6)] at which the population spatial distribution becomes Poisson, and the community heterogeneity critical threshold (CHCT), at which the metacommunitys becomes homogenous. The three concepts—NHCT, CHCT, and PACD—share the same mathematical formula [*i.e.*, Eqn. (6)], but they differ in interpretation, corresponding to distinct ecological organizations: network, metacommunity, and population.

A key difference between the network heterogeneity measured with TPLoN parameters and the heterogeneity of a community (or metacommunity) measured with conventional TPL/TPLE is that the former synthesizes species interaction (or co-occurrence) information beyond mere fluctuations in species abundances. In fact, the focus on measuring network diversity has been increasing in recent years (Ohlmann et al. 2019, Luna et al. 2020, Ma & Li 2024). Network diversity typically also measures the diversity of species interactions. Therefore, both the fields of network heterogeneity and diversity are interconnected and share certain aspects.

## 7.3. Scale free networks and network science

Barabasi & Albert (1999) initially discovered that the degree distribution, or the vertex connectedness, of numerous complex networks adheres to a scale-free power law distribution (Box S1). They proposed that the fundamental mechanism generating this power-law distribution is the preferential attachments to nodes that are already well-connected in a self-organized manner. Nguyen & Tran (2012) contended that the degree isn't the sole crucial factor affecting the network growth of power law networks. Rather, each network node should possess a fitness measure that gauges its ability to attract links. This fitness measure is more comprehensive than degree in generating power law networks. Zhang et al. (2021) explore the applicability of power laws to complex networks. They develop a theoretical network model based on random node addition and deletion, along with preferential attachment. The degree distribution of this model exhibits a clear power law pattern, supporting their commonality in networks. However, the authors note that real networks' finite size and node disappearance probability can obscure power laws. They suggest that detecting power laws effectively involves observing the degree distribution's limiting behavior within its valid interval. The paper also shows the power law exponent depends on the node disappearance probability, explaining why different networks have varying exponents.

The definition of a scale-free network has never been precisely established (Arita 2005). The most commonly cited characteristic of scale-free networks is the topological (scale) invariance of a network structure, which remains consistent regardless of how coarsely it is viewed (Arita 2005). A scale-free network contains highly connected nodes, referred to as hubs, which also includes global hubs (like Google on the Internet), irrespective of the scale at which the networks are observed. This scale-invariance is also known as self-similarity (Box S1), highlighting the self-similar structure whether it is viewed on a global or local scale. Scale-free networks have the further property that they exhibit resilience against random failures or random rewiring, and susceptibility to targeted attacks. However, topology alone doesn't provide enough information to investigate the functions and dynamics of a specific complex system (network). The degree distribution may or may not reflect the critical process underlying network functions and dynamics (Arita 2005).

In one dimension, a power law distribution (Box S1) exhibits a scale-free characteristic. This characteristic becomes apparent on a log-scale, where the slope parameter remains independent of the scaling of the X-axis. However, the scale-free property isn't necessarily connected to network topology, as the power law distribution itself doesn't inherently pertain to networks. Only when the power law distribution is applied to the degree distribution does it become related to networks, particularly scale-free networks. Yet, a power law degree distribution alone doesn't ensure a scale-free network, as it's possible to randomly swap the links without changing the overall degree distribution. Moreover, a scale-dissimilar network may also possess a power law degree distribution.

The reasons behind the widespread occurrence of the power law remain unknown. The fascination with power law distributions dates back to the 19th century. Power law distributions have been discovered or rediscovered in various fields, including the number theory in mathematics (Newcomb in 1881, and Frank Benford in 1938) (Newcomb 1881; Benford 1938), economics (Pareto distribution in 1896) (Pareto 1896), Zipf's law in linguistics in 1949 (Zipf 1949), Taylor's power law in 1961 (Taylor 1961), Barabási-Albert (1999) in physics, and Yule in 1925 (Yule 1925).

The distribution of nodes with degree $k$ adheres to a power law, represented as $P(k) \propto k^{-\alpha}$ (Barabasi & Albert 1999). However, Broido & Clauset (2019) in their recent research suggest that the strongly scale-free network structure is not commonly observed empirically. They evaluated 927 complex empirical networks and concluded that in the majority of cases, the log-normal distribution provides a fit to the network data that is as good as, or better than, the power law. They classified scale-free networks into five categories, ranging from weakest to strongest, and found that while 57% of the networks fall into one of these categories, only 4% are part of the strongest category. Certain biological and technological networks fall into the highest level, while social networks are, at best, only weakly scale-free.

A limitation of the scale-free theory is that it, along with most complex network properties, assumes that the networks are infinite. Hence, the application of scale-free network methods to finite networks can hardly be precise. Holme (2019) contended that the determination of whether a distribution is heavy-tailed is much more critical than whether the power law distribution passes the test. Holme (2019) stated: "Still, it often feels like the topic of scale-free networks transcends science—debating them probably has some dimension of collective soul searching as our field slowly gravitates toward data science, away from complexity science (Box S1)."

## 7.4. Information theory, graph theory and network science

Freitas et al. (2019) showcased the integration of information theory and complex networks. Specifically, they established network entropy using Shannon entropy and Fisher network information measure for complex networks. The network entropy is a property of the network as a whole, while in contrast, the Fisher information measure is a local property as it is highly sensitive to local disturbances. Their research is a component of the so-called entropy-complexity plane, which comprises entropy measures ($H$) and statistical complexity ($C$). To define and calculate statistical complexity ($C$), two pieces of information are required, namely, the information content and disequilibrium. The former can be quantified with entropy and the latter can be assessed with the divergence between the current state of the system and a suitable reference state.

Drawing inspiration from foundational studies in sociology (social networks) and economics (Borgatti et al. 2009, Jackson 2010, Easley et al. 2010), complex networks initially borrowed its primary concepts from graph theory, specifically the groundbreaking work on random graphs by Erdos and Renyi (1961) (Box S1, Watts & Strogatz 1998, Barabasi & Albert 1999). However, since that time, the fields of network science and graph theory have seen limited cross-fertilization and overlap.

Iñiguez et al. (2020) advocated an increased cross-fertilization between graph theory and network science. They argued that, despite graph theory being viewed as the bedrock of network science due to the early links between the evolution of random graphs and network science, both fields largely operate independently today. Their commentary was quite comprehensive regarding bridging the division (gap) between graph theory and network science. Specifically, they highlighted potential advancements in understanding the role of randomness in modeling, hinting at underlying behavioral laws, and predicting complex networked phenomena in nature that have not yet been observed. They contended that "any framework disregarding heterogeneity of elements or interactions will fail in accurately describing and predicting the behavior of complex interacting systems". Network science aims to depict both components and interactions with the goal of modeling the structure and dynamics of complex systems.

At present, graph theory is primarily concerned with providing rigorous proofs for graph properties, including graph enumeration, coloring, and covering, with broad applications spanning from chemistry to circuit design. Conversely, network science bears more resemblance to phenomenological physics, with its focus on observing real-world networks and employing ad hoc mathematical/statistical concepts for their measurement, with the aim of understanding their underlying generative mechanisms. For instance, graph theory has been centered on structures that are more amenable to analysis, such as random or dense graphs, while network science has been concentrated on the most prevalent characteristics of big data, such as sparsity and heterogeneities in the structure and dynamics of large but finite networks (Iñiguez et al. 2020).

The scientists of DYNASNET (dynamics and structure of networks)[1] are aiming to link the concepts of interest in graph theory (e.g., the representative sets and limits objects) to the big ideas of network science, including degree heterogeneity, structural and dynamic correlations, communities structures and mesoscopic order, and coupling and synchronization between the dynamics of and on complex networks Iñiguez *et al.* 2020). Real-world networks of physical, biological, social and technological systems are spatially heterogeneous at several scales, with mesoscopic patterns of core-periphery structures. Furthermore, real-world networks are also temporally dynamic in the sense that network nodes and links may appear and disappear governed by heterogeneous dynamics, not necessarily deterministic (Iñiguez *et al.* 2020). Iñiguez *et al.* (2020) suggested that advances in graph theory, including limit objects for stochastic block model networks (Karrer & Newman 2011), the temporal networks (Holme 2012), and random matrix theory for sparse graphs (Mehta 2004), and control theory for control network (Yan *et al.* 2017) are promising in offering solutions for dealing with the previous challenges posed by real-world networks. Bridging the gaps between graph theory and network science, which will also enhance the similarity between mathematics and reality in complex systems and in turn allows scientists to forecast events, control natural systems, and even predict behavior that has not yet been confirmed empirically (Iñiguez *et al.* 2020).

Xiao *et al.* (2021) formulated the node-based fractal dimension (NFD) and the node-based multifractal analysis (NMFA) framework to identify the generative rules and quantify the scale-dependent topological multi-fractal characteristics of a dynamic complex network. This includes new indicators for gauging the complexity, heterogeneity, and asymmetry of network structures, as well as the structural distance between networks. Their framework can assist in pinpointing the phase transitions in networked systems and comprehending the multiple generative mechanisms underlying the evolution of the network (Xiao *et al.* 2021). Although TPL was not involved in Xiao et al. (2021) framework, developing similar framework with TPL is highly feasible.

---

[1](https://www.ceu.edu/dynasnetproject)

## 8. TPL and Other Classic Power Laws in Macrobial Ecology

This section provides an overview of other classic power laws that are frequently intertwined with TPL in macrobial ecology. These laws are particularly prevalent in the context of species diversity (Box S1), body size, metabolic rate, and population density. Reuman et al. (2008) highlight four allometric power law scaling relations in community food webs, including individual size distribution (ISD) and species-mean-size distribution (SMSD), both of which use a power-law probability distribution. They also introduce the local size-density relationship (LSDR), a power law function linking population densities and mean body masses of species. Additionally, they propose the concept of Variance-Mass Allometry (VMA), a power law relationship between the variance of population density and mean individual body mass, which was validated using data from oak trees (Cohen et al. 2012; Cohen et al. (2016) further examine the relationship between TPL and associated allometric power laws across space, time, and environment, using datasets from mountain beech forests in New Zealand. Another power law in community ecology, the power-function relationship between metabolic rate and body mass, is analyzed by Hudson et al. (2013) using a comprehensive database of animal measurements. Ma (2018a, 2018b, 2019) extended classic species-area relationship (SAR) to general diversity-area relationship (DAR), and further integrated it with TPL. He & Gaston (2003) investigate the link between the variance-mean relationship and the occupancy-abundance relationship in ecology, proposing a unified model connecting species abundance, spatial variance, and occupancy. Lastly, Cohen (2020) studies the relationship between species-abundance distributions (SADs) and TPL of fluctuation scaling (TL), focusing on MacArthur's broken-stick SAD model.

TPL is not the sole power law observed in ecology. In fact, several other power laws, notably those related to species diversity, body size (mass), metabolic rate, and population density, are particularly prevalent, in addition to TPL. While TPL was initially mainly examined in the realm of population ecology, the other ecological power laws have been extensively explored in community ecology, particularly in the context of food webs.

In relation to the connection between population density and body mass (size), there exist at least four allometric power law scaling relations in community food webs, as indicated by Reuman et al. (2008). Two categories of allometric power laws are typically formulated, focusing either on individuals or species (taxon). When concentrating on individuals, the individual size distribution (ISD) is a probability density function employed to describe the distribution of individual-organism body masses, disregarding their taxonomic identities. The ISD can be viewed as a transformed abundance spectrum, essentially capturing the same information in a different form. When focusing on species, the species-mean-size distribution (SMSD) examines the distribution of species-mean body masses using a probability distribution model, similar to ISD but utilizing mean species body mass. Both ISD and SMSD employ a power-law probability distribution, that is, a univariate probability distribution model, akin to the model used in modeling the degree distribution of scale-free networks. A third allometric power law is a bivariate power law function, similar to TPL. Also focusing on the species level, the local size-density relationship (LSDR) refers to a power law function between population densities and mean body masses of species. This third type of allometric power law investigates the scaling of population densities (N) against mean body masses (M) of species, in the power function, $N = aM^b$. The approximate scaling exponents of ¾ and 2/3 have been widely reported, but are based on two different hypotheses, respectively (Reuman et al. 2008). Reuman et al. (2008) utilized theories and data from 146 community webs to study the forms (models) of the three allometric relations and their interrelationships.

While TPL relates the variance of population density with the mean of population density, the density-mass allometry (DMA) links the mean population density with the mean individual mass; both connections are in power functions (Cohen *et al*. 2012). Cohen *et al*. (2012) hypothesized that both power laws, namely TPL and DMA, could be interconnected to generate a direct power law relationship between the variance of population density and mean individual body mass. They termed this new power law relationship as Variance-Mass Allometry (VMA). They validated the VMA power law using data from oak trees (Cohen *et al*. 2012).

Cohen et al. (2016) further explored the relationship between TPL and associated allometric power laws across space, time, and environment, using datasets from mountain beech forests in New Zealand. Conceptually, this is the same integration as in Cohen et al. (2012) of TPL (variance of population density vs. mean population density) and the allometric power law of mean population density vs. mean biomass per individual, which results in a new power law of population density and mean biomass per individual. In other words, the mean population density serves as a link to couple both power laws.

A fourth power law in community ecology is the power-function relationship between metabolic rate and body mass ($Metabolic\ Rate = aM^b$). Perhaps due to the challenges in data collection, most studies on the metabolic rate-body mass power law have been based on basal or resting metabolic rates, and/or have utilized data of species-averaged masses and metabolic rates. Conversely, studies using field, individual metabolic rates are relatively scarce. It's clear that the latter type of datasets is more realistic and more directly reflect ecological interactions and, therefore, natural selection (Hudson et al. 2013). Hudson et al. (2013) compiled and analyzed a comprehensive database comprising measurements of 1498 animals from 133 species in 28 orders. They found the mean scaling exponent to be 0.71 and 0.64 for birds and mammals, respectively. It's suggested that the scaling exponent of 2/3 is governed by the "surface law of metabolism"—a theory based on the ratio of volume to surface area, which influences the rates of heat production and dissipation to the environment. Conversely, the scaling exponent of ¾ is governed by Kleiber's law—a theory based on the scaling of circulatory systems and other biological networks. They also discovered that (i) taxonomic heterogeneity in scaling exponent is statistically significant and substantial relative to the difference (3/4-2/3) (gap between the two theories), and also relative to the difference between the slopes of birds (0.71) and mammals (0.64); (ii) variation is most important at the order and species levels.

A fifth power law in ecology is the classic species-area relationship (SAR). The SAR was discovered in the 19th century, describes the correlation between the number of species or species richness (S) and the area (A) where these species are found, using the power function *i.e.*,

$$S = cA^z. \tag{29}$$

This relationship was expanded into a more comprehensive diversity-area relationship (DAR) by incorporating Hill numbers (Hill 1973) as diversity measures (Ma 2018). The equation for this is:

$$^qD = cA^z, \tag{30}$$

where *q* represents the diversity order (*q*=0, 1, 2, …), and *c* and *z* are fitted parameters.

Before Ma (2018a, 2018b, 2019), there were attempts to extend the classic SAR using alternative diversity metrics, such as Shannon entropy. However, these efforts yielded inconsistent results. The successful expansion of SAR to DAR can be attributed to the unique property of Hill numbers, which are the *effective* number of species or species equivalents. Other diversity metrics, like Shannon entropy (Box S1), typically do not possess this property, even though many of these metrics are functions of Hill numbers and can be converted from one to another. Ma (2019) further extended this concept to the diversity-time relationship (DTR) (Ma 2018b) and the diversity-time-area relationship (DTAR) (Ma 2019).

Additionally, Ma (2018a, 2018b, 2019) modified the power function by introducing a third parameter (*d*) or an exponential decay term exp(*dA*), resulting in the power law with exponential cutoff (PLEC), represented by

$$^qD = cA^z \exp(dA). \tag{31}$$

Using the PLEC model, the maximum accrued diversity (MAD) across a metacommunity was derived as follows (Ma 2018a, 2019b, 2019):

$$Max(^qD) = {}^qD_{max} = c(-\tfrac{z}{d})^z \exp(-z) = cA_{max}^z \exp(-z) \tag{32}$$

where $A_{max}$ is the number of areas accumulated to reach the maximum and is equal to A,

$$A_{max} = -z/d. \tag{33}$$

This maximal accrual diversity, also known as potential diversity or dark diversity, includes species that are present locally and those that are absent locally but present in the regional species pool. Coupling TPL and DAR-PLEC models can provide confidence interval estimates for diversity, with TPL being utilized to estimate the variance of the diversity estimate (Ma 2021, 2022). This approach has successfully estimated global COVID mortality with a precision exceeding 90% (Ma 2021, 2022).

He & Gaston (2003) investigate the link between the variance-mean relationship and the occupancy-abundance relationship in ecology. Using eight empirical data sets, they demonstrate that these patterns can predict each other and propose a unified model connecting species abundance, spatial variance, and occupancy. Mechanistically, these patterns can be explained by metapopulation dynamics and the "rescue effect," which increases occupancy and decreases spatial variance. The study underscores the importance of studying spatial variation for understanding metapopulation and species persistence. The occupancy-abundance relationship, which relates the proportion of areas occupied by a species to its average abundance, offers insights into metapopulation dynamics and species' range limitations.

Cohen (2020: Theoretical Ecology) investigates the relationship between species-abundance distributions (SADs) and Taylor's power law of fluctuation scaling (TL). The author focuses on MacArthur's broken-stick SAD model, which represents relative species abundances as spacings along a partitioned line segment. Cohen demonstrates that this model asymptotically obeys TL with an exponent of 2, meaning the variance of species abundance equals the square of the mean species abundance. This relationship, confirmed numerically for communities with as few as 19 species, suggests a connection between SADs and variance functions. While other SAD models like the lognormal distribution (Box S1) can also exhibit power laws, the broken-stick model provides a clear analytical example. The study not only establishes a link between these two key areas of ecological theory but also opens new avenues for investigation in theoretical ecology, raising questions about how other SADs relate to power law or other variance functions. This connection has broader implications as both models have applications in various scientific fields.

Arguably, the most commonly used models in biodiversity research are based on information entropy (Box S1). However, the paper "The Marine Diversity Spectrum" by Reuman et al. (2014) appears to be an exception. They developed a mechanistic model is developed to understand how diversity varies with body mass in marine ecosystems, leading to the prediction of the 'diversity spectrum'. They investigated the distributions of species body sizes within both taxonomic groups and geographical assemblages, providing insights into various ecological and evolutionary processes. The marine diversity spectrum is predicted to be approximately linear across a mass range spanning seven orders of magnitude on the log-scale, which manifests the power-law scaling patterns. Their study predicts and validates different power-law scaling slopes for the global marine diversity spectrum, which represent different community structures. The relationship between diversity and body mass is explained through the dependence of predation behavior, dispersal, and life history on body mass, and a neutral assumption about speciation and extinction (Reuman *et al.* 2014).

# 9. TPL in Microbiome Ecology

This section examines TPL in microbiome ecology through seven thematic subsections. The first subsection traces foundational work, including experimental validation of TPL in bacterial microcosms (Kaltz et al., 2012) and theoretical extensions to community-level patterns (Ma, 2012a, 2015, Ma & Ellison 2024, Ma 2025). Subsequent subsections explore: human microbiome applications, from postmortem microbial dynamics (Pechal et al., 2018) to spatial organization in gut microbiota (Riva *et al.,* 2019) and core microbiome concepts (Neu et al., 2021); host genetic influences on microbial distributions (Shoemaker, 2022; Liu et al., 2022); TPL as a macroecological framework for microbial spatial ecology (Wetherington et al., 2022) and community complexity (Galbraith & Convertino, 2021); diagnostic applications analyzing gut microbiota stability in health and disease (Martí et al., 2017; Li et al., 2019); connections between TPL and other power laws, including microbial island biogeography and rRNA gene scaling relationships (Kaufman et al., 2017; Gonzalez-de-Salceda & Garcia-Pichel, 2021); and finally, the Binary Power Law (BPL) (Hughes & Madden, 1992)—a special case of Taylor's Power Law (TPL)—reformulates the variance-mean (*V-M*) relationship as a variance-binomial (*V*-$V_B$) function, where $V_B$ represents the variance of a binomial random variable (e.g., disease incidence). Given its structure, the BPL offers distinct advantages in fields like plant pathology, disease ecology, and medical ecology.

## 9.1. TPL for microbial populations and TPL extensions to community ecology

Kaltz *et al.* (2012) experimentally confirmed Taylor's Power Law (TPL) for microbial populations within a bacterial microcosm environment and examined the impacts of abiotic and biotic stress as well as genetics on TPL parameters. They found that both evolutionary and ecological factors can collectively affect TPL parameters. While in most instances, the *b*-value fell between 1 and 2, the highest *b*-value they recorded was *b*=5.43, which is uncommon in the TPL literature.

The experimental method used by Kaltz *et al*. (2012) to validate TPL closely aligns with previous field observations in macrobial ecology, involving the sampling of plant and animal populations. However, the rapid expansion of TPL applications can be attributed to the revolutionary metagenomic technologies used to study microbial genes, particularly with the initiation of the Human Microbiome Project (HMP). This project produced extensive metagenomic datasets, leading to the creation of operational taxonomic unit (OTU) tables, which can be seen as the equivalent of species abundance distribution (SAD) data in macrobial ecology. Against this backdrop, Ma (2012a, 2015) suggested expanding TPL to the community level, introducing four TPL extensions (TPLE).

TPL was discovered in studies of biological populations, particularly insects. Much of the early research on TPL was confined to population ecology, with a few exceptions such as Taylor & Woiwod (1982) who applied TPL to multi-species data, Shaw & Dobson (1995) who applied TPL to mixed species ensemble. It's worth noting that Iwao (1977) also proposed the concept of interspecies mean crowding. However, due to a lack of multispecies datasets, neither TPL nor mean crowding was systematically investigated at the community scale. The advent of metagenomic sequencing technologies and the launch of the HMP (Human Microbiome Project) led to an accumulation of extensive microbiome data. This provided the background for Ma (2012a) to propose extending TPL to the community scale by leveraging the rapidly accumulating microbiome data since the 2010s. Ma (2015) defined four TPL extensions (TPLEs): Type-I TPLE for measuring community spatial heterogeneity; Type-II for community temporal stability; Type-III TPLE for mixed-species spatial aggregation (Box S1); and Type-IV TPLE for mixed-species temporal stability.

The so-termed "mixed-species" concept is similar to the multi-species setting in the early stage of TPL and mean crowding studies (Taylor & Woiwod 1982, Iwao 1977). The term aggregation (Box S1) is used since 'mixed-species' is essentially an 'average' species level entity and population aggregation (Box S1) can be considered as the counterpart of community heterogeneity. Ma's (2015) TPLEs use the same power function form as the classic TPL, but the mean (M) and variance (V) of TPLEs are defined differently from those in the classic TPL. Accordingly, the interpretations for the TPLE parameters are different from those in the classic TPL. For example, for Type-I TPLE, the mean (M) is defined as the mean species abundance (*i.e.*, per species) in a community, and V is the corresponding variance. The Type-I TPLE is built by regression V-M series across a series of communities (*i.e.*, metacommunity scale). Therefore, the parameters of Type-I TPLE measure the heterogeneity at community scale, strictly, metacommunity scale, which is of essential difference with the classic TPL. Similarly, Type-II TPLE measures the community (temporal) stability at metacommunity scale. Furthermore, Ma (2015) also extended the concept of population aggregation critical density (PACD) to community scale as community heterogeneity critical threshold (CHCT), at which point a community is considered homogenous, and to network heterogeneity critical threshold (NHCT) (Ma 2025).

## 9.2. TPL in Human Microbiomes

Pechal *et al.* (2018) analyzed how taxon abundance and variability in the postmortem microbiome changed over time. They found this relationship followed TPL, which describes variation in ecological systems. The slope indicated less variation within two days of death. TPL typically has a slope of two, but ecological systems often have a lower slope. The results suggested the microbiome was stable in the first two days after death, becoming less stable later on potentially due to competitive interactions driving succession. Riva *et al.* (2019) utilized TPL to analyze the spatial heterogeneity of the colon microbiota. They suggested that dietary fiber and polysaccharides can influence the spatial organization of the colon microbiota.

Neu *et al.* (2021) delve into the concept of the core microbiome, noting the absence of standardized methods for its definition and quantification. They analyze various approaches, including taxon occurrence rates and relative abundances, and highlight the influence of factors like sampling scale, sequencing depth, and taxonomic level on core microbiome results. The paper also discusses the relevance of abundance-occupancy relationships, consistent with TPL, in constraining the composition of the core microbiome. TPL suggests a positive correlation between the local abundance of a species and the number of sites it occupies. This relationship is particularly relevant when defining and quantifying the core microbiome, as it can help identify microbial taxa that are consistently present across samples.

Marti *et al.* (2018) applies TPL to taxon abundance data from multiple time points in a gut virome study. By fitting scaling models to these data, the researchers could quantify and correlate variance-mean scaling patterns between different bacterial and viral taxa. Analyzing how well these taxa conform to TPL and how their scaling patterns correlate provides insights into potential coevolutionary selection. Cohen *et al.* (2017) aimed to establish a theoretical framework linking the distribution of parasites within hosts to their spatial distribution. They collected field data on parasite and host populations in New Zealand lakes and found that TPL described the relationship between mean and variance for host and parasite abundance per square meter.

## 9.3. Influences of host genetics and evolution on microbiomes

Shoemaker et al. (2022) studied the influence of human genetics on the gut microbiome. Utilizing sparse canonical correlation analysis (sCCA), the study identifies associations between human single nucleotide polymorphisms (SNPs) and the composition and function of the gut microbiome. The findings highlight that both common and rare microbial functions are linked to host genetics, offering a novel framework for investigating host-microbiome genetic interactions. Methodologically, they employed the Tweedie distribution to model the taxa and gene abundances derived from metagenomic shotgun sequencing, owing to its adaptability in handling right-skewed, over-dispersed, and zero-inflated data. The Tweedie distribution, characterized by a large mass at zero and a long positive tail, is determined by its shape parameter, $p$. When $1 < p < 2$, the distribution is continuous for $Y > 0$, making it suitable for zero-inflated data. The variance of the response variable is tied to the mean by the Tweedie power and dispersion parameters, $p$ and $\varphi$, as per the equation $Var(Y) = \varphi \mu^p$. The study found that the Tweedie distribution, particularly when $p$ falls between 1 and 2, effectively captures the relationship between the mean and variance in the abundances of metagenomic taxa and genes, outperforming the negative binomial and Poisson distributions. It also accurately accounts for the number of zeros in the data, with the predicted number of zeros closely matching the observed count. This is modelled as:

$$Y(Y = 0) = \exp\left[-\mu^2 - \frac{p}{\varphi(2-p)}\right], \tag{34}$$

where φ is the dispersion parameter and $p$ is the power parameter, and the likelihood of a gene count in a sample being zero decreases linearly as the mean increases.

Shoemaker *et al.* (2022) discusses the potential of comparative population genetics in understanding the human gut microbiome, an area yet to be fully explored. It suggests that examining patterns across species can reveal evolutionary trends, such as purifying selection dominating as lineages diverge. The reanalysis of existing data revealed pathways like glycolysis facing stronger purifying selection, while specialized pathways have relaxed constraints. The study concludes that this approach can identify new genes and evolutionary patterns, and pinpoint exceptional species and genes. Shoemaker et al. (2022) also explores the d*N*/d*S* ratio, a metric in population genetics measuring natural selection on DNA sequences. It represents the rate of nonsynonymous (amino acid-changing) to synonymous (neutral) substitutions. A ratio *<1* indicates purifying selection, *1* suggests neutral evolution, and *>1* indicates positive selection.

The study finds a slight increase in the variance of d$N$/d$S$ faster than the square of the mean d$N$/d$S$ across pathways, consistent with Taylor's power law, suggesting an underlying power law scaling between the variance and mean of d$N$/d$S$ across metabolic pathways in gut microbiome species.

Liu et al. (2022) investigated amphibian skin microbes' distribution patterns and their correlation with host body size. Using 358 specimens from 10 species, they applied Ma's (2015) Type-I and III Taylor's power law expansions (TPLE) to evaluate microbial community heterogeneity. The study found consistent, highly aggregated microbial abundance, suggesting a shared heterogeneity scaling parameter, indicative of scale invariance, across different hosts. Sarabeev *et al.* (2017, 2022) introduces a macroecological framework for analyzing host-parasite relationships in invasive species, using metrics like prevalence, abundance, and aggregation. TPL is a core component of the macroecological framework proposed in this paper. TPL-*b* value is used to characterize aggregation patterns and help distinguish co-introduced *vs*. acquired parasite groups. It applies this framework to helminth parasites in native and invasive grey mullet fish populations.

### 9.4. TPL for understanding microbiome distribution and dynamics

Wetherington et al. (2022) study the influence of habitat landscapes' spatial structure on microbial population distribution, specifically *Escherichia coli* in microfluidic landscapes. They assess the impact of corridor width variance on bacterial distributions, revealing that higher variance changes population occupancy fluctuations in line with TPL. This suggests that beyond a certain variance level, the landscape's spatial structure significantly impacts long-term microbial distribution.

Galbraith & Convertino (2021) analyzed the complexity of bacterioplankton communities in various marine habitats, characterizing population variability through TPL. This law describes the scaling relationship between a population's variance and mean abundance, represented by the exponent ν. At the community level, ν ranged from 1.5 to 1.7, aligning with the *b*-parameter of TPL. Interestingly, they found an inverse linear relationship between ν and the power law exponent (ε) used to characterize the abundance distribution, but a weaker relationship with the exponential distribution parameter (λ) used for species interactions. This suggests ν and ε may encode similar community structure signals, while λ measures a different variation aspect. TPL offered insights into population stability within communities and across habitats.

Björk et al. (2018) examined temporal variability and mean abundance patterns within microbiomes. They found core and transient taxa to be more stable and abundant than opportunistic taxa. Using TPL, they observed an exponent less than two, suggesting variability (heterogeneity) decreases with increasing abundance. This indicates weak species interactions can reduce fluctuations of common versus rare taxa. The study demonstrates that intraspecific interactions and environmental stochasticity contribute to stabilizing core microbiome dynamics, as predicted by TPL.

Ji et al. (2020) highlights that TPL accurately characterizes the relationship between the average abundance and temporal variance of gut microbiota species. It demonstrates that temporal abundance variances scale as a power law function of average abundances across bacterial taxa, with exponents ranging from 1.49-1.71. Notably, Taylor's law also facilitated the identification of specific bacterial taxa displaying abnormal temporal dynamics during dysbiosis events. Taxa associated with travel and infection exhibited over 10-fold greater variance than expected by TPL, identifying them as outliers. This underscores the utility of Taylor's law as a quantitative marker for perturbations in gut microbiome ecology.

Ji et al. (2019) developed DIVERS (Decomposition of Variance using Replicate Sampling), a quantitative framework that systematically disentangles three fundamental sources of variation in bacterial community data. The model distinguishes between temporal dynamics (changes in microbial abundances over time), spatial sampling variability (differences across physical sampling locations), and technical noise (variation introduced during experimental processing). In their analysis of both human gut and soil microbiomes, temporal dynamics capture ecological fluctuations, spatial variability reflects habitat heterogeneity, and technical noise accounts for measurement artifacts inherent to laboratory and sequencing procedures.

Wang & Liu (2020) explore the origins of various scaling laws in microbial dynamics, including Taylor's law and new scaling laws related to the degree distribution of the Visibility Graph. Their method converts a microbial time series (*e.g.*, species abundance or metabolic rate over time) into a network (graph). Each data point in the time series becomes a *node* in the graph. An *edge* connects two nodes if they are 'visible' to each other, meaning no intermediate data point obstructs the line between them. These scaling laws are universal across different habitats, from human skin and oral microbiome to marine plankton bacteria and eukarya communities. Using the Generalized Lotka-Volterra (GLV) population dynamics model, the authors confirm that these scaling laws are determined by inter-species interactions and linear multiplicative noises, largely driven by temporal stochasticity of the host or environmental factors.

Ho et al.'s (2022) findings highlighted the significance of power law relationships in understanding microbial community dynamics. Specifically, the researchers found that time series data from various microbiotas exhibited power law scaling between a taxon's abundance variance and its mean over time, *i.e.*, the TPL. Furthermore, the distributions of residence and return times, which represent periods of sustained presence or absence of a taxon, also followed TPL. This suggests that the dynamics of microbial communities are characterized by long periods of dominance by a few taxa, punctuated by rapid shifts in community composition. They suggested that the observed power law patterns could be driven by resource competition among taxa and their responses to environmental fluctuations.

George & O'Dwyer (2022) presents a unified framework, the stochastic linear response model (SLRM), to analyze population dynamics across diverse systems like microbial communities, forests, and cities. The model, with three parameters, can reproduce empirical distributions of abundances and fluctuations across these systems. The model predicts the distribution of population fold-changes over time, providing a null model for quantifying large fluctuation risks. The Equilibrium to Noise Ratio (ENR), a key SLRM parameter, is approximately conserved across different microbial species. This conservation leads to TPL scaling, suggesting universal properties underlying fluctuations.

Zamponi *et al.* (2022) investigates if large microbial ecosystems exhibit critical dynamics properties. The researchers analyzed human microbiome and plankton community datasets over time, and showed variance scaling super-linearly with mean abundance, consistent with TPL and indicative of critical dynamics. This suggests long-range correlations between taxa over time, a hallmark of critical phenomena. The study concludes that microbial ecosystem dynamics operate near a critical point, offering both high variability and robustness.

Zamponi et al. (2022) explore whether large microbial ecosystems display critical dynamic properties. They analyze human microbiome and plankton community datasets, focusing on temporal fluctuations in microbial taxa abundance. Their findings reveal power law distributions in abundances and changes in abundance, suggesting scale-free fluctuations. The data demonstrated the variance scaling with mean abundance on log-scale, aligning with TPL and indicative of critical dynamics. This implies long-range temporal correlations between taxa, a characteristic of critical phenomena. The study concludes that microbial ecosystem dynamics operate near a critical point, providing both high variability and robustness. The scaling of taxon abundance fluctuations over time, as per Taylor's law, provided crucial evidence.

Zaoli and Grilli (2021) analyze long-term gut microbiome data from multiple individuals to characterize microbial abundance variability. They find most microbial taxa display stationary fluctuations around a constant carrying capacity, consistent with the stochastic logistic model (SLM). The SLM predicts that noise intensity, $\sigma$, should follow Taylor's law and be independent of the carrying capacity, $K$, across taxa. The empirical estimates confirm this prediction. Some taxa display abrupt transitions between alternative carrying capacities, suggesting alternative stable states. Differences between individuals are largely explained by differences in carrying capacities and independent stochastic fluctuations. The paper shows a macroecological framework, centered around the SLM and TPL, can statistically describe both intra- and inter-individual variability.

Shoemaker et al. (2022) investigates the genetic diversity in the human gut microbiome across individuals and species. It identifies invariant patterns of genetic diversity, including a relationship between mean allele frequency and variance following TPL. The paper applies the Stochastic Logistic Model (SLM) to these patterns, predicting a gamma distribution of allele frequencies. The SLM's predictions align with independent estimates of strain structure, suggesting strain-level ecology primarily governs detectable genetic variation in the gut microbiome.  Similarly, Wolff et al (2023) applied SLM and TPL to explore the diversity-stability relationship within the human gut microbiome at both the species and strain levels. Their findings suggest that the macroecological properties of the human gut microbiome, including its stability, emerge at the level of strains, highlighting the importance of strains as a unit of ecological organization in the human gut microbiome.

Shetty et al. (2022) explores the complex interactions within the human gut microbiome, influenced by host-derived glycans and diet. To understand these interactions, a synthetic minimal microbiome, named Mucin and Diet based Minimal Microbiome (MDb-MM), was designed using 16 key bacteria. The research used various methods to examine community dynamics, stability, and metabolic interactions. The study found that these 16 species co-existed in in vitro gut ecosystems containing a mixture of complex substrates. The MDb-MM followed Taylor's power law, demonstrating similar ecological and metabolic patterns. The microbiome showed resistance and resilience (Box S1) to temporal changes, as seen in the abundance and metabolic end products. The study concluded that microbe-specific temporal dynamics in transcriptional niche overlap and trophic interaction network explain the observed co-existence in a competitive minimal microbiome. This research provides valuable insights into the co-existence, metabolic niches, and roles of key intestinal microbes in a dynamic and competitive ecosystem (Shetty et al. 2022).

## 9.5. TPL for disease diagnosis

The Human Microbiome Project (HMP) sought to define core microbial communities across individuals but confronted profound compositional heterogeneity shaped by host genetics and environment (HMP Consortium, 2012). This variability prompted a paradigm shift - from taxonomic inventories to functional network stability, where conserved metabolic pathways outweigh specific microbial signatures (Lloyd-Price *et al.*, 2017). Notably, disease associations with microbial diversity (Box S1) or heterogeneity show complex patterns: only ~30% of cases demonstrate significant diversity/heterogeneity-disease relationships (DDR/HDR), with distinct mechanistic bases underlying these associations (Ma *et al.,* 2019; Ma, 2020b). The heterogeneity-disease relationship (HDR) was analyzed with TPLE approach (Ma, 2020b). The HMP's legacy thus reframed heterogeneity as a biological feature rather than noise in microbiome-host interactions.

Martí et al. (2017) analyzed 16S rRNA gene sequencing and shotgun metagenomic sequencing data from 99 subjects to study temporal fluctuations in gut microbiota compositions. They found that the relative abundances of taxa in gut microbiota over time followed TPL. The parameters from TPL, variability (V) and scaling index (α), indicated microbiota stability. Healthy subjects had lower V values, suggesting more stability over time. The researchers modelled temporal dynamics using the Langevin equation, identifying stable and unstable phases. Subjects' health status correlated with their microbiota stability, with unhealthy subjects often in the unstable phase. The study suggests that a healthy adult microbiota is generally stable over time, while disruptions can lead to instability and potential dysbiosis.

Džunková et al. (2018) examined the temporal variability of oral microbiomes in 26 volunteers over a 30-day period. They used a power law to model changes in the relative abundance of operational taxonomic units, employing two parameters: α (equivalent with TPL-*a*), which represents the amplitude of fluctuations and overall stability, and β (equivalent with TPL-*b*), which describes the statistical behavior. This model effectively captured individual-level dynamic changes in microbiome composition over time. However, their power law model utilized *dispersion* instead of variance, a common practice in analyzing temporal fluctuation with TPL.

Li & Convertino (2019) investigates the dynamics of the human gut microbiome in healthy and irritable bowel syndrome (IBS) individuals using network inference models. It reveals that healthy networks display a balanced, scale-free topology, while unhealthy networks appear more random. The study identifies an inverse relationship between species' total outgoing information flow and relative species abundance (RSA). Notably, it demonstrates that TPL has different scaling exponents for healthy and unhealthy groups. This links RSA fluctuations to network topology and stability. Despite the unhealthy group having higher α- and γ-diversity, the healthy group shows optimal diversity growth rate and RSA variability. The healthy microbiome adheres to TPL with lower RSA variability, reflecting a more organized, resilient network. Conversely, the unhealthy microbiome exhibits larger RSA fluctuations, tied to its more random topology. Therefore, Taylor's law scaling provides insights into the contrasting stability of healthy and dysbiotic microbiome states. Nevertheless, in a comprehensive heterogeneity-disease relationship (HDR) study, it was found that the HDR is significant in less than 1/3 of the cases studied (Ma 2020b).

## 9.6. Other power laws in microbiome research

Kaufman et al. (2017) explore the application of MacArthur and Wilson's theory of island biogeography, which explains biodiversity patterns across isolated geographic areas, to the microbial scale. Large-scale genomic analyses reveal that the cumulative number of unique genes discovered increases as a power law function of the number of genomes sampled. This was demonstrated for *Salmonella, E. coli,* and *Campylobacter.* The power law exponents measured suggest scale-invariant behaviour extends to microbes and microbial genes.

Gonzalez-de-Salceda and Garcia-Pichel (2021) explored the relationship between microbial cell size and the number of ribosomal RNA gene copies per cell. They compiled data on over 100 species of bacteria, archaea, and microbial eukaryotes, examining cell volume and ribosomal gene copies per cell. They found a power law relationship, with the number of ribosomal gene copies scaling with an exponent of 2/3 relative to cell volume. This contrasts with total DNA content per cell, which scales with an exponent closer to 3/4. The authors suggest this allometric scaling relationship allows for a simple correction to account for biases in characterizing microbiome community structure and taxon-specific abundances using ribosomal gene sequencing. The reason behind the 2/3 scaling relationship remains unknown but may relate to cellular protein needs. This mathematical rule enables improved quantitative microbiome analyses across microbiology fields. As a related point, the Diversity-Area Relationship (DAR) Power Law (PL) and Power-Law with Exponential Cutoff (PLEC), which were discussed in Section 8, can also be applied to microbiomes. Indeed, the extensions from the Species-Area Relationship (SAR) to DAR have been tested with human microbiomes (Ma 2018, 2019). The reason for their previous introduction is due to the fact that the classic SAR was primarily developed within the context of macrobial ecology of plants and animals.

## 9.7 Binary power law

Hughes & Madden (1992) binary power law (BPL) can be considered as an elegant recasting of TPL by replacing the mean ($M$) in the TPL equation with the variance that the population would have if it were Poisson distributed ($V_p$) (Madden et al. 1995, 2018; Taylor 2022) This led to the formulation of a relationship between the observed variance ($V$) and the variance for a random set generated by the binomial distribution. The BPL is of the following form:

$$\ln(V) = \ln(a) + b\ln(M) = \ln(a) + b\ln(V_p). \tag{35}$$

The above substitution is valid because, for Poisson distribution, $V=M$. Given that the Poisson distribution is a limiting case of the binomial distribution when the probability is small and the sample size is large, the variance of the binomial distribution ($V_B$) can be similarly utilized to compare the observed variance of data distributed binomially.

$$\ln(V) = \ln(a) + b\ln(M) = \ln(a) + b\ln(V_B). \tag{36}$$

The binomial distribution can be naturally applied to model the presence/absence (incidence) data such as the healthy or diseased samples in plant pathology and similarly human diseases. Furthermore, $V_B$ can be estimated with the probability of disease incidence ($P$). Obviously, BPL is a special case of TPL, but for some applications such as disease diagnosis, BPL can be more convenient.

## 10. Perspectives

In retrospect, the dual-pronged explorations led by mathematicians, statisticians, and other scientists (primarily ecologists and entomologists) have woven a 'punctuated' landscape, drawing an analogy with the concept of a punctuated fitness landscape. The eight themes – population spatial aggregation and ecological mechanisms, TPL and skewed statistical distribution, mathematical/statistical mechanisms (generation) of TPL, sample vs. population TPL, population stability, synchrony, early warning signals for tipping points, TPL on complex networks, TPL in macrobiomes, and TPL in microbiomes – along with the timeline of three periods (punctuations) and the future, can be mapped onto this landscape, which is abstracted as Fig 1. The three punctuations along the timeline and the two prongs along the themes divide the landscape into eight regions, each covered by one of the eight themes.

The somewhat modest beginnings of TPL might give the impression that it is a rather simple model, possibly with straightforward interpretations for its mechanisms. However, one observation has made is that, although the TPL model itself is indeed simple, the phenomena it successfully describes are far from simple. In fact, most are complex, and the underlying mechanisms can be even more intricate than initially perceived. This complexity is demonstrated by the expansion of TPL research from simple insect populations to human gut microbiomes and fluctuations in stock liquidity in the 'physical' prong of exploration. In the theoretical prong, universality (Box S1) is a property of power laws, including TPL, and TPL has been identified in everything from the Tweedie distribution to the fitness distribution in computational intelligence and the distribution of prime numbers.

Thanks to the dual-pronged exploration, many of the debates surrounding TPL have been naturally resolved by now. For instance, the early debate on whether there is a single ecological mechanism to interpret TPL has become a moot point, given that TPL has been found to govern the V-M relationship across a variety of complex systems. It is unlikely that a single mechanism can explain all phenomena governed by TPL. Similarly, the question of whether an observed TPL is a statistical artifact can also be misleading, as TPL is still a mathematical model of the complex system or phenomenon. If the V-M relationship is not interesting from an anthropocentric view, one might argue that a perfectly fitted TPL is a 'statistical artifact.' In some cases, there may not be an obvious interpretation for V-M; the distribution of prime numbers (specifically the deviations from their predicted positions based on Riemann's counting formula) is one such example (Kendal & Jørgensen 2015).

In perspective, we propose three potential avenues for future research in Taylor's Power Law (TPL). It is important to note that these suggestions are not exhaustive and are likely biased, as they are inevitably influenced by our current knowledge and perspective. The first direction we suggest is purely directional, while the other two are thematic.

A potential direction involves fostering a reciprocal interaction between the two prongs of TPL. The physical prong could pose new challenges for theoreticians, primarily mathematicians and statisticians, within the abstract prong. In turn, the abstract prong could offer guidance or insights for the effective application of TPL. This interaction is intended to be bidirectional. For example, TPL has been employed to identify early warning signals for tipping points. Other potential applications could involve using population synchrony research for managing natural resources, such as fisheries, or for monitoring the spillover of zoonotic diseases from wildlife to humans. Furthermore, ecologists and evolutionary biologists might pose questions to mathematicians and statisticians, such as whether there are any differences between discrete quantities like counts of individuals or genes and continuous quantities like mutation rates or selection pressures, assuming both types of quantities adhere to TPL. Another example could be, assuming the fluctuation of connected neurons over time follows TPL, how might the scaling parameter influence information processing in a neural network? This question is particularly relevant for understanding and designing neural network-based deep learning algorithms.

The second direction for TPL research could involve using TPL as a model to measure heterogeneity (more precisely, heterogeneity scaling), particularly within the context of complex networks (Ma 2025). Heterogeneity (Box S1) is a concept that is often conflated with diversity, especially in ecological discussions of biodiversity. However, there are fundamental differences between heterogeneity and diversity. Shavit and Ellison (2021) distinguished between diversity and heterogeneity by suggesting that a zoo is diverse, while an ecosystem is heterogeneous. In essence, when assessing diversity in a zoo, one simply tallies the variety of animals present, without anticipating interactions between different species, which are separated by fences or other barriers. Conversely, when evaluating heterogeneity, both interspecies interactions and the relationships between species and their heterogeneous habitats should be taken into account.

While there have been challenges in the past, such as an overabundance of diversity metrics for practitioners to choose from, measuring diversity has become relatively straightforward, with information entropy (Box S1) often providing a sufficient solution. Specifically, the introduction (Hill 1973) and reintroduction (Chao et al. 2014) of Hill numbers for measuring biodiversity have unified many biodiversity metrics (Ricotta & Feoli 2024). However, measuring heterogeneity has proven to be more elusive (Box S1). To the best of our knowledge, there is not yet a widely recognized heterogeneity metric or metric system that holds a similar position to Hill numbers in measuring diversity. This is likely due to the significantly higher complexity of heterogeneity compared to diversity, as heterogeneity must account for interactions or relationships among groups of objects, such as species or individuals.

In the existing body of heterogeneity research, one category of studies often employs diversity metrics (e.g., Hill 1973, Chao et al. 2014) to measure heterogeneity. For instance, the field of tumor heterogeneity extensively borrows from ecology, using diversity indices to measure tumor heterogeneity (Kashyap et al. 2022). However, unless diversity metrics are used to quantify interactions, as done by Luna et al. (2021) and Ohlmann et al. (2019), what these metrics measure is largely diversity, not heterogeneity.

In the existing literature on heterogeneity (Box S1) research, another category of studies often uses statistical variance as a foundation for measuring heterogeneity. For instance, in statistics, particularly in meta-analysis, variance is frequently used as a basis for measuring heterogeneity (Higgins & Thompson 2002), and there's even a related concept called heteroskedasticity (or heteroscedasticity) associated with it (Dutilleul & Legendre 1993). In social sciences, especially econometrics, heterogeneity in initial endowments and the concept known as the "hand of the past" (or state dependence) are often foundational (Heckman 1991, 2001). If the initial endowments have a temporarily lasting influence on outcomes, then heterogeneity is present. If the effects of initial endowments are attenuated or accentuated by subsequent experiences of the process being studied or by related processes, then the 'hand of the past' is at play. As a result, complex mathematical models are used to quantify the heterogeneity associated with unobserved or unobservable variables in econometrics (Heckman 2001), and variance or variance partition often play significant roles in modeling heterogeneity. Similarly, the variance-mean based TPL has been extended to quantify community/landscape heterogeneity as demonstrated by Ma (2015) and Ma & Ellison (2024). However, there is an issue with using TPL to measure heterogeneity, as what traditional TPL essentially measures is the consequence of heterogeneity (e.g., population or community species abundances). To directly deal with interactions or relationships, adapting TPL to complex networks, such as the previously discussed TPLoN (Ma 2024) that models the V-M series of species connectedness directly, should be more advantageous than traditional TPL, as in this case, species interactions are directly incorporated into the TPL.

A third direction we suggest for TPL research is to explore TPL within the context of evolution, particularly with temporal data. Evolution fundamentally involves temporal changes, which are typically gradual, especially in the context of Darwinian evolution. However, in the context of macroevolution, events often appear to occur in a non-gradual manner, following a pattern known as punctuated equilibrium. This pattern involves long periods of evolutionary stability interrupted by rapid shifts between states (Bakhtin et al., 2021). Ma (2012b, 2013) demonstrated that in evolutionary computing, the fitness of optimization follows TPL. Ma & Ellison (2024) suggested, through big data analysis of animal gut microbiomes, that microbial diversity and heterogeneity scaling parameters from power law models (diversity-area power-law relationship for diversity, and TPL for heterogeneity scaling) measure different aspects of the microbial community and/or landscape, given that both scaling parameters are independent (uncorrelated). They further found that a continuous evolutionary timeline (ET) can provide a more powerful lens than phylogenetic taxa (punctuated equilibriums) in detecting the influences of evolution on diversity and heterogeneity. Ma & Ellison (2024) further postulated that in the case of animal gut microbiome, heterogeneity is primarily shaped by ultimate evolutionary characteristics such as evolved dispersal behavior and interspecific interactions, while diversity is primarily shaped by the fluctuations of species abundances on proximate ecological time scales. In conclusion, integrating evolutionary principles (both continuous gradualism and punctuated equilibriums) into TPL research could prove to be a fruitful avenue.

## Standard Declarations

**Acknowledgements**

This review work was generously supported by a Charles Bullard Fellowship from Harvard University, the National Science Foundation of China (NSFC #72274192), and Prosperous Yunnan Talent Program.

**Author Contributions**

MZS conceived the review and wrote the manuscript; RAJT revised the manuscript. Both authors approved the submission.

**Conflict of Interests**

The authors declare no conflicts of interests.

**Data/Code Availability**

Not applicable.

**Online Supplementary Information: Text Box S1 (Glossary)**

# Online Supplementary Information (OSI) for: Six Decades Post-Discovery of Taylor's Power Law: From Ecological and Statistical Universality, Through Prime Number Distributions and Tipping-Point Signals, to Heterogeneity and Stability of Complex Networks

**Text Box S1.** Glossary of Important Terms Supplementing This Review

| No. | Terms | Interpretations |
|---|---|---|
| [1] | *Aggregation* | Population *aggregation* in population ecology can be considered as the counterpart of community *heterogeneity* in community ecology. In population ecology, terms such as dispersion, patchiness, and population spatial distribution are commonly used interchangeably with aggregation. Taylor's Power Law (TPL) was initially developed in the context of studying population spatial aggregation, particularly with a focus on insect populations (Taylor 1961, 1984, 1986). |
| [2] | *Black-Box Approach* | The *black-box* approach intentionally disregards the internal workings of a system, often due to their unknown nature, and instead concentrates on analyzing the output signal from the system. The appropriateness of the black-box method is usually determined by the specific problem and the context in which it is being studied. For example, in the field of biomedicine, it may be ethically inappropriate to observe the microscopic interactions between elements such as cells and external factors like radiation that causes tumors. Fascinatingly, by studying the characteristics of signals, such as scale invariance and universality, we may be able to gain some insight into the nature of the underlying process within the complex system. |
| [3] | *Complexity science* | *Complexity science*, also known as *complex systems science*, is a relatively new subject that studies how relationships between parts give rise to the collective behaviors of a system and how the system interacts and forms relationships with its environment. It is inherently interdisciplinary because it seeks to understand the underlying patterns and structures that emerge across different domains, from biology and physics to social systems and economics (Font-Clos 2015).<br><br>Complexity science often involves the use of computational models and simulations to understand these patterns, and it often heavily borrows concepts, principles, and methodologies from statistical mechanics and critical phenomena to tackle complex systems or issues that can be abstracted from virtually any natural or social science fields (Font-Clos 2015). It focuses on systems that are dynamic, richly interactive, and have emergent properties - that is, properties that are not apparent from the components alone. Examples of complex systems include, but not limited to the human brain, human microbiome, ecosystems, the Earth's climate, economic markets, and the Internet. |
| [4] | *Diversity* | Diversity and heterogeneity, while often conflated, are distinct concepts. Both describe variations or differences among groups of objects, yet they diverge in their consideration of interactions among groups or objects. As Shavit & Ellison (2021) eloquently put it: a zoo exemplifies diversity, while an ecosystem represents heterogeneity. To quantify diversity, a zookeeper simply counts the number of animal species housed in separate enclosures. However, to assess heterogeneity, one must take into account the inter-species interactions, and even the interactions with species environment. A heterogeneous system must be diverse but a diverse system is not necessarily heterogeneous, and therefore, heterogeneity can be a special case of diversity. Hill numbers (Hill 1973), derived from Renyi entropy, provide a comprehensive series of diversity numbers (metrics), encompassing total species number (species richness), the count of representative species, and the |

| | | number of dominant species, among others. The Hill numbers offer robust solutions for quantifying diversity, especially given that virtually all other existing diversity metrics are found to be functions of Hill numbers (Chao et al. 2014, Ricotta & Feoli 2024). However, measuring heterogeneity presents a greater challenge due to the necessity of accounting for interactions. Currently, there is no universally accepted metric for heterogeneity. Many of the existing heterogeneity metrics are tied to statistical moments, particularly variance and skewness. The measurement of heterogeneity is a crucial aspect of Taylor's Power Law (TPL), as highlighted in the works of Taylor (1984, 1986), Ma & Ellison (2018, 2019), Ma & Taylor (2021) and Ma & Ellison (2024), Ma & Li (2024). |
|---|---|---|
| [5] | *Early warning signal (EWS) for Tipping Points* | Early warning signals (EWS) for tipping points (TP), also known as critical transitions in complex systems, can vary depending on the context. However, in general, there are several common indicators: (1) Critical Slowing Down (CSD): As a system approaches a tipping point, it tends to recover more slowly from perturbations. This can be measured by an increase in autocorrelation (the correlation of a signal with a delayed copy of itself) or an increase in variance. (2) Increased Spatial Correlation: As a system nears a tipping point, patterns of spatial correlation can change, often increasing. (3) Flickering: When a system is close to a tipping point, it may start to 'flicker' between states. This can be seen as sudden, temporary shifts to an alternate state before returning to the original state. (4) Skewness: As a system approaches a tipping point, its distribution may become skewed. Also see the entry for '*tipping points*.' TPL is found to reveal EWS for many complex systems, perhaps due to its power law nature. |
| [6] | *Entropy & Information Theory* | Entropy is a concept in physics and information theory that measures the randomness or disorder within a set of data. Specifically, Renyi's entropy is a generalization of the traditional entropy concept. It is a one-parameter family of measures of uncertainty, randomness, and complexity based on the probabilities of outcomes. This measure is particularly useful in various fields such as quantum information, signal processing, and machine learning. Shannon entropy is still the most widely used metric for measuring biodiversity in ecology. Renyi's entropy was introduced to ecology by Hill (1973) but did not receive the attention it deserves until Chao *et al*. (2014) reintroduced it to biodiversity research (*see* Diversity entry in this box).<br><br>Information theory, a branch of applied mathematics and electrical engineering, quantifies information, measuring the efficiency of data transmission and the degree of uncertainty in a dataset. Information theory provides the mathematical framework to measure and describe system complexity, while complexity science applies these tools to understand the properties and behaviors of complex systems. |
| [7] | *1/f Noise* | 1/*f* Noise, also known as pink noise or flicker noise, is a signal or process with a frequency spectrum that falls off at high frequencies with a power law exponent of 1. It's called "pink noise" because light with this spectral graph appears pink. This type of noise is common in electronic devices and can be found in a wide range of systems in nature, including biological systems, traffic flow, and even music and art. It's characterized by the fact that every octave carries an equal amount of noise power. The name 1/*f* comes from the noise's power spectral density (PSD), which is inversely proportional to the frequency of the signal.<br><br>Both TPL (Taylor's power law) and 1/*f* noise are examples of scaling laws, which describe how the characteristics of a system change with its size. They are both observed in a wide range of natural and social systems, suggesting some common underlying principles. |
| [8] | *Graph Theory, Complex Networks & Complexity Science* | Graph theory is a branch of mathematics that studies graphs, which are mathematical structures used to model pairwise relations between objects. A graph in this context refers to a collection of vertices or 'nodes' and a set of |

|   |   | edges that connect these nodes. Graph theory provides a fundamental framework to represent and analyze data in a variety of fields, including computer science, biology, and social sciences. It allows for the examination of relationships between entities and can be used to solve problems related to network connectivity, path optimization, scheduling, and many others.<br><br>In the context of complexity science, graph theory plays a vital role. Complexity science often deals with systems that can be modeled as complex networks, where nodes represent individual components and edges represent interactions between them. These could be neurons in a brain, species in an ecosystem, or computers in a network. Graph theory provides the mathematical tools to analyze these networks, helping to understand the emergent properties, behaviors, and the overall complexity of the system. This makes it a crucial component in the study of complex systems. *See* Iñiguez *et al.* (2020) for detailed introduction, and also *See* 'scale-free network' entry. |
|---|---|---|
| [9] | *Heterogeneity* | *Heterogeneity* can be defined as the differences in groups of system objects that interact with each other and often also with their environments. Heterogeneity is a key feature of complex systems that contributes to their emergent properties, adaptability, robustness, and overall complexity, specifically: (1) Emergent Properties: Heterogeneity can lead to emergent properties, which are characteristics of the whole system that are not present in any of the individual parts. These properties arise from the interactions between diverse components. (2) Adaptability: Heterogeneous systems are often more adaptable to changes in their environment. Because they contain diverse, often redundant components, they have a wider range of responses to any given change. (3) Robustness: Heterogeneity can contribute to the robustness of a system, or its ability to maintain function in the face of disturbances. If one component fails, others with different attributes might still be able to carry on due to redundancy. (4) Complexity: Heterogeneity can increase the complexity of a system, making it more interesting and challenging to study. It can lead to non-linear dynamics, where small changes can have large effects, and vice versa.<br><br>Heterogeneity and diversity are often conflated with each other, but with fundamental differences (Shavit & Ellison 2021); *see* previous 'diversity' entry. |
| [10] | *Lognormal Distribution and Power Law Distribution* | A lognormal distribution is a probability distribution of a random variable whose logarithm is normally distributed. It is often the result of multiplicative processes, where the product of many independent, identically distributed and positive random variables leads to a lognormal distribution. Lognormal distributions are often used to model skewed data, where the values are not symmetrically distributed around the mean, but rather have a long tail on the right side.<br><br>Power law distribution, on the other hand, is a functional relationship between two quantities, where a relative change in one quantity results in a proportional relative change in the other quantity, independent of the initial size of those quantities. In other words, one quantity varies as a power of another. Power law distributions often emerge from lognormal distributions under certain conditions, such as non-uniform data sampling times or when a lower boundary is established during a random walk. A key characteristic of power law distributions is their "heavy tails", meaning they contain more large values than would be expected in a lognormal distribution. This results in power law distributions having the potential for infinite variance, while lognormal distributions have finite variance. However, in practical applications with finite data sets, infinite variance is |
| | | not possible. *See* Stumpf & Porter (2012) for a review on power laws. |
| [11] | *Phase transitions and critical phenomena* | According to Font-Clos (2015), *phase transitions* and *critical phenomena* serve as archetypal instances of universality. For instance, the shifts between the three states of water (ice, vapor, and liquid) represent typical phase transitions, contingent on critical temperatures ($T_c$=0°C or 100°C). Another example pertains to magnetization and temperature. The overall magnetization (M) of a ferromagnetic material decreases as temperature (T) increases. There exists a critical temperature, beyond which the magnetization falls to zero. As the temperature nears this critical temperature Tc, the magnetization follows a power law of the distance to $T_c$, $M \propto |T - T_c|^\beta$, where β is known as the scaling exponent. Interestingly, many materials exhibit the same scaling exponent, and only a few different scaling exponent values have been identified after extensive experimental testing on numerous materials. In the context of a macroscopic material system, composed of numerous elements (the physical particles that form the material), the interaction of these elements (particles) through their local magnetization produces a macroscopic feature of the system—the global magnetization. This behavior near the critical point, as characterized by the power law function and scaling exponents, is independent of the microscopic specifics of the system—the particular composition of the material. Hence, phase transitions in ferromagnetic materials exemplify a typical universality feature in material systems, as stated by Font-Clos (2015). |
| [12] | *Scale-invariance or Self-Similarity* | Intuitively, *scale-invariance* means that objects do not change when they are "zoomed" in or out. The "zoom" may be in spatial coordinates such as microscope or telescope, time, frequency, or other coordinate systems such as taxonomic units. The objects can be a pure mathematical object such as Cantor's set, a physic concept such as pink noise, or something discovered in nature such as coastlines, magnitudes of earthquakes, spatial distribution of populations (Font-Clos 2015).<br>According to Font-Clos (2015), the concept of scale-invariance, also referred to as self-similarity, scale-free, or absence of characteristic scale, implies that an object remains unchanged regardless of whether we zoom in or out. Without additional information, it's impossible to determine the size of what one is observing. In the extreme case of perfect scale-invariance, no concept of typical size (position) can be established. The scale-invariant functions in one dimension are represented by power laws, *i.e.,* $f(x) = ax^b$. Generally, the set of functions *f* that satisfy the scale-invariance property in one dimension is:<br>$$\lambda_0 f(x) = f(\lambda x); \quad \lambda_0, \lambda \in R^+; x \in R.$$<br>It can be verified that power law function satisfies the above condition for scale-invariance since $\lambda_0 ax^b = a(\lambda x)^b, \lambda_0 = \lambda^b$. Parameter *b* is the scaling exponent: it means that if one rescales *x*-axis by a factor of *λ*, then the *y*-axis must be rescaled by a factor of $\lambda^b$ for the power law function to be invariant. |
| [13] | *Scare free networks* | Barabasi & Albert (1999) initially discovered that the degree distribution, or the vertex connectivity, of numerous complex networks adheres to a scale-free power law distribution. They proposed that the fundamental mechanism generating this power-law distribution is the preferential attachments to nodes that are already well-connected in a self-organized manner (Holme 2019). The scale-free network has never been defined precisely (Arita 2005). The commonly stated property of scale-free networks is the topological (scale) invariance of a network structure, regardless how coarsely it is viewed (Arita 2005). A scale-free network must have highly connected nodes, known as hubs, which must include global hubs (such as the Google on the Internet), regardless of the scale of the observed networks. The scale-invariance is also known as self-similar, which emphasizes the self-similar structure regardless |

| | | of whether it is viewed globally or locally. |
|---|---|---|
| [14] | *Stability, Resilience & Persistence* | Th  Stability, in the context of systems theory, refers to the ability of a system to return to its equilibrium state after a disturbance. It's a measure of how much a system can be perturbed before it transitions to a different state. Resilience and persistence are closely related to stability. Resilience is the capacity of a system to absorb disturbances and still retain its basic function and structure. It's about how much a system can adapt to changes and disturbances without changing its fundamental nature. On the other hand, persistence refers to the ability of a system to continue over a long period, often despite challenging conditions.<br><br>In essence, stability, resilience, and persistence are all measures of a system's ability to withstand disturbances. Stability is about the system's ability to return to an equilibrium state, resilience is about its ability to adapt and maintain its basic function, and persistence is about its ability to endure over time. These concepts are often used together to understand and predict the behavior of complex systems in the face of disturbances.<br>It's important to note that the terms stability, resilience, and persistence have numerous definitions, with over a hundred in ecology alone. Similar terms, such as reliability, survivability, and sustainability, have been adopted in other disciplines (Ma 2008). Reliability and survivability are extensively studied in engineering and computer science, while sustainability is popular in environmental science and public policy. These terms are often associated with concepts like black swan events, tipping points, and grey rhino events. These concepts extend into a wide array of fields, such as operations research, survival analysis, game theory, econometrics, management sciences, sociology, and political and military sciences. Indeed, power laws are frequently discovered to provide a fitting representation of the data or patterns scrutinized within these disciplines. However, the underlying mechanisms driving these observations often remain unclear, posing one of the greatest challenges in the application of power laws. |
| [15] | *Stochasticity & Randomness* | Randomness and stochasticity are two closely related concepts that both deal with unpredictability and the lack of a discernible pattern. Randomness, in its broadest sense, refers to the absence of any predictable order in events. It is a fundamental concept in various fields such as mathematics, physics, philosophy, and computer science, where it denotes the occurrence of events without a deterministic reason but with a certain statistical distribution. Stochasticity, on the other hand, is often used to describe systems or processes that are inherently unpredictable due to the influence of random variables. While it shares the element of unpredictability with randomness, stochasticity often implies a degree of statistical predictability. In a stochastic process, the probability distribution of the outcomes is known, even though the individual outcomes themselves are unpredictable. Thus, while both concepts revolve around unpredictability, stochasticity often carries an additional layer of statistical understanding. |
| [16] | *Synchrony* | Synchrony, also known as spatial synchrony or metapopulation synchrony, is the propensity for time series of population densities, observed in various spatial locations, to correlate over time. Synchrony operates within the metapopulation framework, though in practice, strictly defining the distributions or boundaries of a local population can be difficult. A particularly intriguing aspect of synchrony theory is the so-called "Moran effect," which refers to synchrony induced by simultaneous environmental drivers. For instance, it is broadly speculated and actively researched whether climate change has been modifying population synchrony in general. See Cohen (2013, 2014a, 2014b), Cohen and Saitoh (2016), and Reuman *et al.* (2017) for more detailed information. |
| [17] | *Tipping point* | A tipping point is a critical threshold in a system at which a minor alteration, such as a slight increase in global temperatures, a small shift in market trends, or a change in species population, can trigger a major transformation. This |
| | | major transformation could be a drastic shift in climate patterns, a sudden crash or boom in an economy, or a significant change in an ecosystem. These changes are often significant, irreversible, and result in a new state or behavior of the system. |
| [18] | *Universality* | According to Font-Clos (2015), *Universality* refers to universal macroscopic laws of nature emerged from a variety of different microscopic dynamics. A universal macroscopic feature of a system is one that is independent of its microscopic behaviors. However, macroscopic observables are usually defined from the microscopic elements, for example, by defining the community diversity with Shannon entropy or by defining the community heterogeneity with Taylor's power law (TPL). A key point in understanding the *independence* is that the independence is only *approximate* or *almost*. That is, the global, universal feature depends only on a few parameters of its microscopic elements, but independent of the rest majority. Some of the statements (such as the law of large numbers) on universal features are exact for *infinite* number of elements ($N$), and good approximations for large but finite $N$ (Font-Clos 2015). |